# Light-Matter Coupling in Scalable Van der Waals Superlattices

Pawan Kumar[1,2], Jason Lynch[1], Baokun Song[1$], Haonan Ling[3], Francisco Barrera[1], Huiqin Zhang[1], Surendra B. Anantharaman[1], Jagrit Digani[3], Haoyue Zhu[4], Tanushree H. Choudhury[4], Clifford McAleese[5], Xiaochen Wang[5], Ben R. Conran[5], Oliver Whear,[5] Michael J. Motala[6], Michael Snure[7], Christopher Muratore[8], Joan M. Redwing[4], Nicholas R. Glavin[6], Eric A. Stach[2], Artur R. Davoyan[3], Deep Jariwala[1*]

[1]Electrical and Systems Engineering, University of Pennsylvania, Philadelphia, PA, 19104, USA
[2]Materials Science and Engineering, University of Pennsylvania, Philadelphia, PA, 19104, USA
[3]Department of Mechanical and Aerospace Engineering, University of California at Los Angeles, CA, 90095, USA
[4]2D Crystal Consortium-Materials Innovation Platform, Materials Research Institute, Pennsylvania State University, University Park, PA, 16802, USA
[5]AIXTRON Ltd., Anderson Road, Swavesey, Cambridge, CB24 4FQ, UK
[6]Air Force Research Laboratory, Materials and Manufacturing Directorate, Wright-Patterson AFB, Dayton, OH, 45433, USA
[7]Air Force Research Laboratory, Sensors Directorate, Wright-Patterson AFB, Ohio 45433, USA
[8]Department of Chemical and Materials Engineering, University of Dayton, Dayton, Ohio, 45469, USA

*Corresponding author: dmj@seas.upenn.edu
$Present Address: Huazhong University of Science and Technology, China

**Abstract:**

Two-dimensional (2D) crystals have renewed opportunities in design and assembly of artificial lattices without the constraints of epitaxy. However, the lack of thickness control in exfoliated van der Waals (vdW) layers prevents realization of repeat units with high fidelity. Recent availability of uniform, wafer-scale samples permits engineering of both electronic and optical dispersions in stacks of disparate 2D layers with multiple repeating units. We present optical dispersion engineering in a superlattice structure comprised of alternating layers of 2D excitonic chalcogenides and dielectric insulators. By carefully designing the unit cell parameters, we demonstrate > 90 % narrowband absorption in < 4 nm active layer excitonic absorber medium at room temperature, concurrently with enhanced photoluminescence in $cm^2$ samples. These superlattices show evidence of strong light-matter coupling and exciton-polariton formation with geometry-tunable coupling constants. Our results demonstrate proof of concept structures with engineered optical properties and pave the way for a broad class of scalable, designer optical metamaterials from atomically-thin layers.

**Introduction**

Semiconducting Multi-quantum Wells (MQWs) and superlattices form the basis of all modern high performance opto-electronic and photonic components, ranging from modulators to lasers and photodetectors[1-4]. However, most known and scalable MQWs and superlattice structures are epitaxially grown III-V, II-VI or oxide perovskites. While significant progress has been made over the last three decades in the commercialization of II-VI and III-V MQWs and superlattices, the inherent difficulties of their integration onto arbitrary substrates have limited their adoption and applicability[5,6]. The advent of van der Waals (vdW) semiconductors, the ability to grow them uniformly over wafer scales, and to transfer them with high fidelity onto arbitrary substrates opens new avenues towards rational design of both electronic and photonic dispersions in artificially stacked superlattices and MQW structures[7].

Two-dimensional (2D) vdW materials are a broad and growing family of materials with a diverse range of electronic properties, encompassing metals, semiconductors, ferromagnets, superconductors and insulators[8-11]. This diversity allows for vdW materials to be combined with one another, or with other thin materials, into heterostructures with new or enhanced properties and improved performance in a variety of applications[7,12]. Most of the previous research has focused on heterostructures that are made using mechanically exfoliated layers that are a few µm$^2$ in lateral size[13] with uneven thickness. This presents significant challenges in making MQW or superlattice structures with sufficient reproducibility across the number of periods necessary to enable the desired photonic or electronic dispersions. Further, with every increasing layer, stacking mechanically exfoliated flakes reduces the size of the region that has the desired overall stacking sequence, a problem that is exacerbated with each additional layer that is added. This fact prohibits the scalability of this approach. For photonic and optoelectronic applications, there is another major challenge with regards to atomically-thin active layers: the nature of light-matter interaction. Monolayer thickness of 2D materials results in a reduced cross section of light-material interaction implying weak couplings, despite naturally resonant optical responses[14].

Semiconducting transition-metal dichalcogenides (TMDCs) that consist of Mo, W, Re etc. are a sub-class of vdW materials that have large, complex refractive indices due to the strong in-plane bonding of the transition metals to the chalcogenides, leading to strong light-matter interactions[15,16]. The low dielectric screening and highly-confined exciton wavefunctions present in TMDCs result in excitonic binding energies of ~500 meV, creating strong excitonic resonances at room temperature[17]. As the thickness of TMDCs decreases from the bulk to the monolayer limit, they transition from indirect to direct bandgap semiconductors. This reduces the non-radiative energy loss of exciton relaxation and leads to an enhancement in photoluminescent (PL) emission[18,19]. However, reducing the thickness to monolayer dimensions adversely impacts the net interaction with light. Therefore, to engineer strong interaction with light and still maintain the key advantages to monolayer scaling it is necessary to make either a metamaterial or superlattice structure with monolayer repeat units in one dimension[20,21].

Here, we report the experimental realization of superlattices that are specifically designed to achieve near-unity absorption while concurrently maintaining the enhanced PL emission and optoelectronic properties of monolayer TMDCs. Our superlattices are cm$^2$ in scale and comprise of repeating unit cells of metal organic chemical vapor deposition (MOCVD) grown TMDCs (MoS$_2$ and WS$_2$) + insulating spacers (h-BN and Al$_2$O$_3$) stacked on an Au back reflector (Figure 1A). We observe the emergence of strongly coupled exciton-polaritons in the superlattices when light is coupled into the superlattices at incident angles > 45°. Further, both the exciton-polariton dispersion and coupling strength are observed to be tunable by altering geometric parameters of the superlattice and its unit cells. This assembly process is both general and simple and allows extreme flexibility regarding materials choice due to the lack of microfabrication constraints. Thus, it can be expanded to allow the integration of several different 2D chalcogenides and spacer layers, as demonstrated here. The resulting approach opens up a broad field of artificially engineered, scalable vdW superlattices for electronic, opto-electronic, and photonic applications.

## Results and Discussion

We adopt a highly scalable approach to the fabrication of vdW MQWs and superlattice structures. Wafer-scale samples of WS$_2$, MoS$_2$ and h-BN grown using the MOCVD technique were used for sample fabrication. Details of sample growth and synthesis are available in the methods section and supporting information. We have adopted and demonstrated two different structures in this work, as shown in **Figure 1a**: 1) An all-vdW superlattice comprised of alternating layers of TMDC and h-BN, and 2) a mixed dimensional superlattice with alternating layers of TMDC and 3D oxides grown via atomic layer deposition (ALD). The all-vdW superlattice predominantly studied in this work is comprised of a unit cell of monolayer $WS_2$ as the excitonic layer with insulating h-BN (3 nm) as a spacer layer which was repeated on top of an $Al_2O_3$/Au substrate. In the TMDC-oxide superlattice, the h-BN layers are replaced by $Al_2O_3$ layers. Our proposed concept and fabrication scheme is quite general and has been extended to monolayers of $WS_2$, $MoS_2$, and few-layer $MoSe_2$ as the excitonic media in this study.

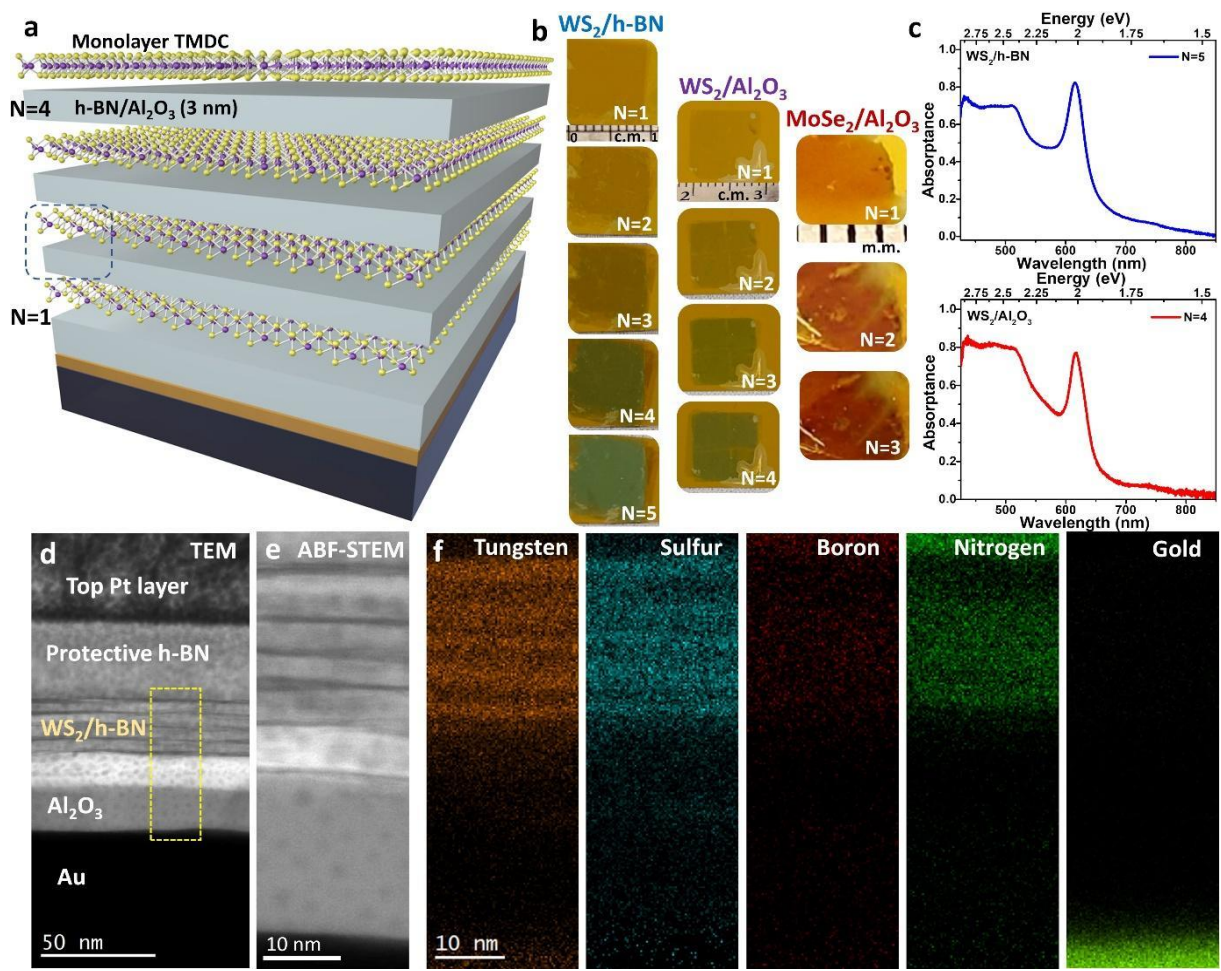

**Figure 1: Structure and Composition of Multilayer Excitonic Quantum Well Superlattices (a) Schematic structure of the multilayer superlattice structure, N=1 and N=4 respectively indicates unit cell number. (b) Optical camera images of ∼ 1x1 $cm^2$ samples of the superlattice with progressively increasing number of unit cells, for monolayer $WS_2$/3 nm h-BN, monolayer $WS_2$/3 nm $Al_2O_3$ and few layer $MoSe_2$/3 nm $Al_2O_3$. The sample color contrast evidently improves with increasing N. (c) Experimentally measured absorptance spectra for $WS_2$/h-BN (N=5) (top) as well as $WS_2$/$Al_2O_3$ (N=4) (bottom) superlattice samples where > 80% absorption at the $WS_2$ excitonic peak is observed. (d) Cross-sectional transmission electron micrographs (bright field), (e) Annular bright field and (f)**

**Energy dispersive x-ray spectroscopy (EDS) maps of the WS$_2$/h-BN: N=5 sample. The electron micrographs and elemental maps show the layered structure clearly.**

We prepare our superlattice structures via wet transfer of the 2D chalcogenide and h-BN layers. In addition, we use atomic layer deposition growth of aluminum oxide for the base spacer layer as well as a component of unit cells, in some cases. These large area superlattice samples require wafer-scale grown TMDCs and h-BN with uniform thicknesses. The MoS$_2$, WS$_2$ and h-BN samples were grown via MOCVD while the MoSe$_2$ samples were grown via selenization of Mo thin films in H$_2$Se gas at elevated temperatures (see methods and supporting info, Figure S1 and S2). The increased absorption of the superlattices as the number of unit cells increases can be seen by the naked eye, as shown in Figure 1b. WS$_2$-based superlattices transition from a near transparent film at N=1 to a deep green color at N=5. The large area (cm$^2$) realization of these superlattices (Figure 1b) is a particularly important demonstration of this study, given that any practical application of 2D TMDCs in photonics/optoelectronics will require large area, uniform material. Experimentally measured absorptance spectra for two-different superlattices have been shown in Figure 1c. A detailed discussion on the absorption properties is in following section.

To examine the spatial uniformity of this multilayer superlattice structure, we perform cross-sectional transmission electron microscopy of our representative sample (WS$_2$/h-BN: N=5) (Figure 1d). An annular bright field (ABF)-STEM image of the cross section of our superlattice sample (Figure 1e) clearly shows the multilayer structure and five dark lines running horizontally in the image contrast, indicating the heavy element layer (WS$_2$). Corresponding energy dispersive X-ray spectroscopic (EDS) elemental maps shows the layers clearly. Layers of W, S and N signals are apparent, as are weak signals for B and Au corresponding to the substrate location (Figure 1f). The white contrast (gap) between the TMDC layers and the bottom alumina (Figure 1d) is a result of delamination of the superlattice during the ion milling process (Xe$^+$ ions) for cross-sectional TEM sample preparation (see supporting info, Figure S3).

A two-variable optimization scheme using Transfer Matrix Method (TMM)-based modelling combined with a genetic algorithm-based optimization approach was used to determine the desired thickness of the superlattice samples. The objective of the optimization was to maximize (minimize) narrow band absorptance (reflectance) at the excitonic resonance. The unit cell spacer and bottom spacer thicknesses were optimized to maximize the excitonic absorptance (see methods and supporting info, Figure S4-S7 for details). We performed this TMM modelling and optimization for four different superlattices (WS$_2$/h-BN, WS$_2$/Al$_2$O$_3$, MoS$_2$/Al$_2$O$_3$ and MoSe$_2$/Al$_2$O$_3$) where the number of unit cells varied from N= 1 to 8. We found a distinct absorptance enhancement up to N=4 and N=5 for Al$_2$O$_3$ and h-BN based superlattices, respectively, as experimentally demonstrated in **Figure 2**.

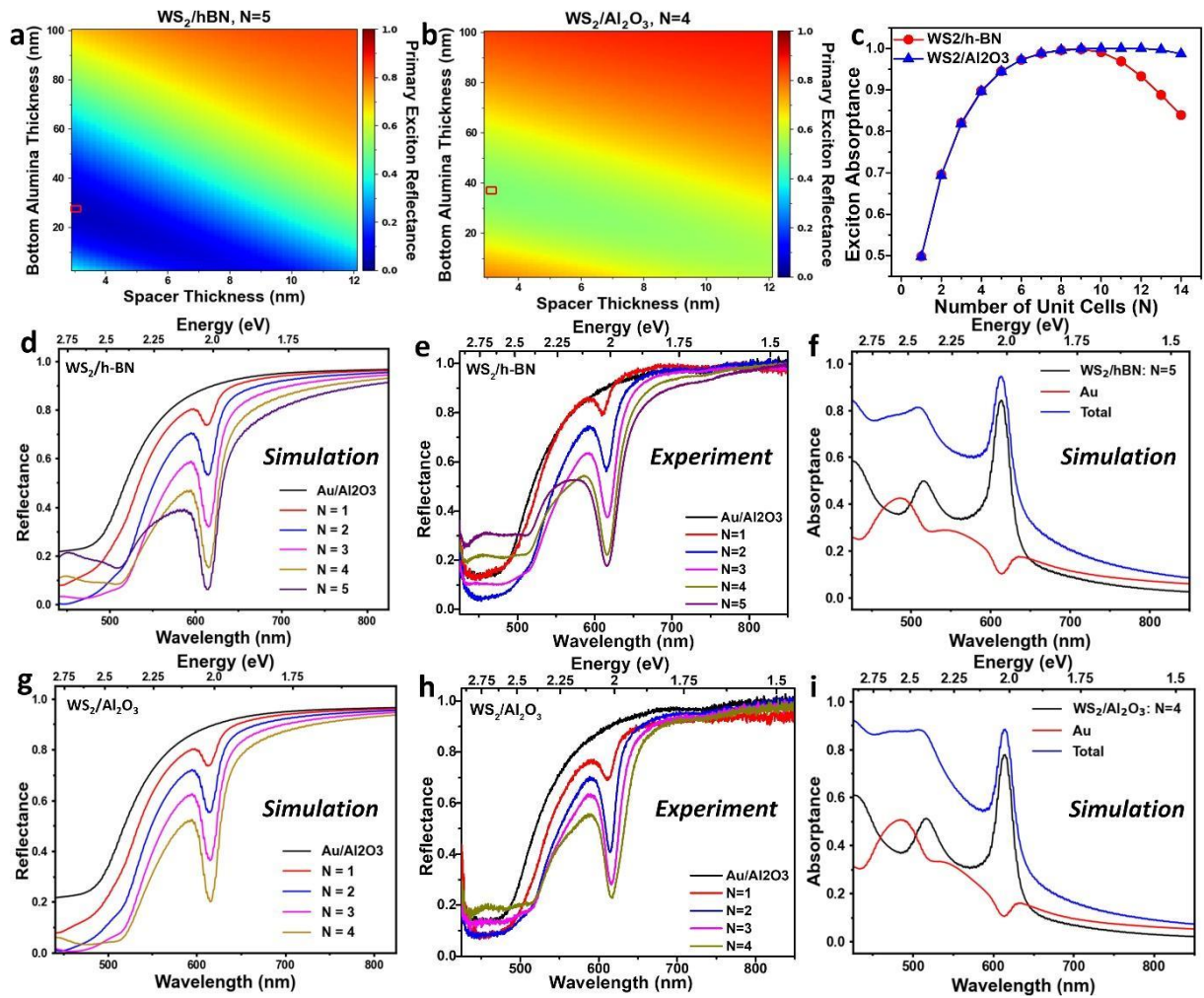

Figure 2: Thickness and Optical Property Optimization. 3D color map of the excitonic reflectance (~ 620 nm) as a function of thickness parameters for the bottom alumina layer and spacer layer, for (a) WS$_2$/h-BN and (b) WS$_2$/Al$_2$O$_3$ samples. (c) The layer dependent, optimized absorptance for WS$_2$/h-BN (N=5), and WS$_2$/Al$_2$O$_3$ (N=4) follows a trend of increasing rapidly before asymptotically approaching a limit. (d, g) Simulated and (e, h) Experimental normal incidence reflectance spectra for a function of increasing number of unit cells in the stack for WS$_2$/h-BN and WS$_2$/Al$_2$O$_3$ superlattices, respectively. The decreasing reflectance with increasing N is evident and shows excellent qualitative and quantitative agreement between simulations and experiments. (f, i) Simulated total absorptance decomposed into absorptance of individual component layers for the corresponding superlattice stacks.

The modelling and optimization identifies three key features in terms of designing superlattice geometric parameters for maximizing (minimizing) absorption (reflection) at the excitonic resonance. They are: 1) the thickness of the spacer layer in the unit cells must be as low as possible (Figure 2a and b); 2) the bottom alumina thickness should be reduced as the number of unit cells (N) increases (see supporting Figure S5-S8); and 3) the peak absorptance (at exciton resonance) approaches near unity with increasing N and then dips again, suggesting that there is an optimal unit cell number (N) for perfect absorption (Figure 2 d-i). We note that this optimal N can vary with spacer index and thickness. Given that electronic interactions between two TMDC layers are non-negligible at spacer thicknesses of 1-1.5 nm[22,23], there is little room for further improvement in light trapping while maintaining electronic isolation. The complex refractive index values used in the TMM simulations were obtained through spectroscopic ellipsometry measurements of the samples that were used for the fabrication of the superlattices. The simulated reflectance of the stack decreases with each successive unit cell deposition at the $WS_2$ excitonic resonance (Figure 2d, g), for both $WS_2$/h-BN and $WS_2/Al_2O_3$ superlattice samples. This is in strong agreement with the experimentally measured spectra (Figure 2e, h). This agreement between simulated and experimentally measured spectra suggests that the dielectric function of $WS_2$ is unperturbed from it's as-grown state on sapphire, when integrated into this multilayer superlattice stack. It also suggests that there is no electronic interaction between neighbouring the $WS_2$ layers when they are separated by h-BN or $Al_2O_3$. Therefore, they maintain their monolayer electronic character when integrated into this superlattice stack. The insulating layers serve two primary purposes: 1) they act as spacer layers between TMDCs to reduce the electronic coupling between individual TMDCs layers, allowing them to maintain monolayer properties and 2) they act as light trapping agents, because the refractive index difference between the TMDCs and insulating layers also results in enhanced reflection. Given the low reflectance (high absorption) at the excitonic resonance, an important parameter to consider is the extent of useful absorption in the semiconductor vs parasitic absorption in the underlying Au. This absorption into individual components of the superlattice was extracted through the TMM simulations (Figure 2 f, i), where a negligible contribution from the bottom Au layer is observed around the primary exciton wavelength (613 nm for $WS_2$) (see supporting info, Figure S8 for additional details). It is further worth noting that there is a pronounced dip in reflectance in the bottom Au at the excitonic resonance. This is a peculiar observation and is attributed to lack of sufficient incident light intensity reaching the Au surface because of multiple reflections and trapping in the layers above. This observation can be generalized to other bottom metals such as Ag, where the parasitic absorptance is further diminished, yet the reflectance dip remains (see supporting info, Figure S9). Likewise, $MoS_2/Al_2O_3$ superlattices also show similar behavior (see supporting Figure S10). We have also analyzed and compared the reflectance behaviour of a multilayer, mechanically exfoliated $WS_2$ control sample with similar thickness to that of the $WS_2$ present in the superlattices ($WS_2/Al_2O_3$: N=4 and $WS_2$/h-BN: N=5) (see supporting info, Figure S11, S12). The comparative analysis confirms that while a similar degree of absorbance can be achieved with equivalent thickness of $WS_2$ on an alumina spacer, there is loss of the direct band gap nature of the TMDC. Additionally, the superlattice exciton peaks remain unshifted, as opposed to the red-shift seen in few-layer $WS_2$ of equivalent absorber thickness. This lack of energy shift further confirms that the $WS_2$ is electronically isolated in the multilayer. Finally, it is worth nothing that the oscillator strength of the exciton is higher in monolayers. This is because of the reduced dielectric screening and quantum confinement, which also results in larger absorption per unit thickness (see supporting info, Figure S12).

The ability of these multilayer superlattices to both trap light and retain their monolayer electronic and optical character is a defining feature of our approach. In contrast to their bulk counterparts, the presence of strong quantum confinement in monolayer $WS_2$ and $MoS_2$ leads to a direct band gap, which in turn leads to high intensity PL due to low, non-radiative energy loss during electron-hole recombination[24]. This extraordinary feature of monolayer TMDCs makes them strong candidates for light emitting devices. The superlattice structures demonstrated herein – which combine insulating spacer layers – between monolayer TMDCs allows the monolayers to maintain their direct-gap electronic structure. We verify this using a series of vibrational and luminescence spectroscopy measurements shown in **Figure 3**.

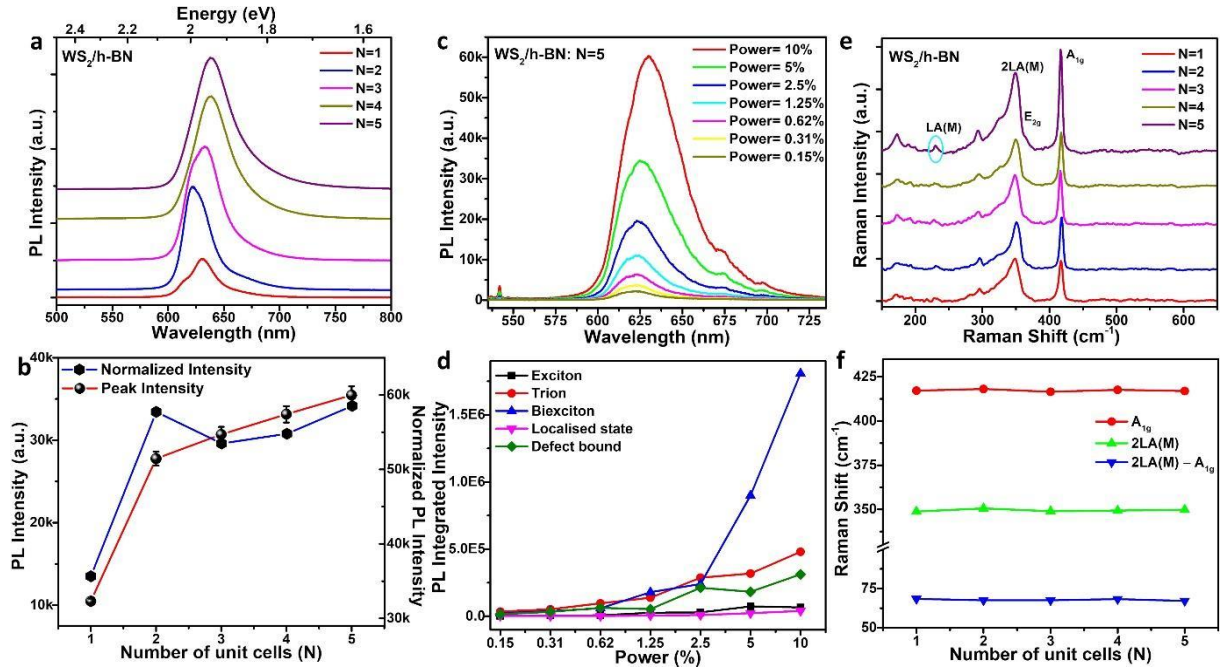

Figure 3: Maintenance of monolayer properties. (a) Room temperature photoluminescence (PL) spectra of the multilayer superlattice as a function of increasing number of unit cells for the $WS_2$/h-BN superlattice. (b) Corresponding PL peak intensity and absorption normalized PL vs. number of unit cells, showing increasing PL intensity with increasing number of unit cells The net PL intensity saturates when the total absorption saturates, while the increase in normalized absorption (at pump wavelength of 405 nm) suggests enhanced light outcoupling by the multilayer superlattice structure. (c) Power-dependent PL spectra for the $WS_2$/h-BN: N=5 superlattice and (d) PL peak intensities emerging from various excitonic components of the power dependent spectra in (c), extracted using Lorentzian fits. The sharp increase in bi-exciton intensity is evident. (e) Raman spectra of the same multilayer superlattice with increasing number of unit cells showing the same peak positions. The peak intensity sharpens and rises with enhanced light interaction for the samples with larger N. (f) Unchanged characteristic Raman peak positions with increasing N.

We observe that our multilayer superlattices remain highly luminescent with an increasing number of unit cells (Figure 3a). Not only are the luminescent properties maintained, but the luminescence intensity is even enhanced with increasing N due to the increased due to enhanced light-material interactions (Figure 3b). When normalized to total useful light absorption with increasing N, using layer resolved absorption calculations we observe that the luminescence intensity increases with N which is expected due to stronger light-material interaction (see supporting info, Figure S13 for additional details). We also observe a small red shift in PL peak positions with increasing number of unit cells (from N=1 to 5) which can be attributed to the increasing density of defect emission or doping arising from trapped contaminants such as PMMA, organic solvents and water that are used in the wet chemical transfer process (see supporting info, Figure S2). Similar PL behavior was observed in other monolayer TMDC based superlattices, namely $WS_2/Al_2O_3$ and $MoS_2/Al_2O_3$ (see supporting info, Figure S14). To further understand any possible effects of enhanced absorption and light trapping on the luminescence characteristics, we have also performed PL characterization as a function of incident pump power. At high powers, emission from various higher order and charged excitonic species are observed. Individual excitonic components (neutral exciton, trion, dark exciton, biexciton, defect bound exciton, or localised state exciton) can be identified by analyzing the power-dependent PL emission of the $WS_2$/h-BN superlattice (N=5) and decomposing the spectra into individual Lorentzian peaks (Figure 3 c, d and supporting info, Figure S15). A notable observation in our power dependent PL analysis is the sharp rise in emission contribution from biexcitons with increasing power. Biexcitons comprise of two-electrons and two-holes and are known to recombine by either decomposing into two excitons or emitting two photons in sequence or a pair entangled photons[25]. Biexcitons tend to form at high excitation levels. In our case the high excitation not only occurs due to increased incident power but also due to enhanced light trapping (~2x). Such cavity-induced enhancement of biexcitons have been reported earlier in perovskite quantum dots and $MoSe_2$ layers[26,27]. The N=1 case (single monolayer) does not show a similar power dependence (see supporting info, Figure S16), further suggesting the that light trapping causes the enhanced excitation in multilayers. Along with increasing biexciton emission, luminescence contributions from defect bound and localised state excitons also appear to increase in the N=5 superlattice. This can again be attributed to defect accumulation during the repeated transfers and thermal cycles of ALD with each increasing N of the superlattice. However, the overall relation between emission intensity vs. excitation stays linear at both room and low temperatures (80 K) within the range of excitation powers probed, thereby suggesting an absence of any non-linear phenomena in this range. While we do observe significant enhancement of emission peak intensity and full width half maxima (FWHM) sharpening of the emission spectra (see supporting Figure S17), there are no obvious signatures of new states or presence of non-linear phenomena.

In addition to PL, Raman spectra provide a strong signature of interlayer interaction and hybridization. Specifically, the out-of-plane vibration mode ($A_{1g}$: 418 cm$^{-1}$) stiffens with increasing number of layers and therefore the separation between 2LA(M) and $A_{1g}$ modes reduces with increasing layer thickness[28]. We observe no noticeable peak shifts in the Raman spectra of WS$_2$ in our superlattices with increasing N. Once again this suggests no interaction between layers and no detectable strain within the layers. The only noticeable difference with increasing N is the rising peak intensity and narrowing FWHM of the peaks (Figure 3e). This is likely due to increased Raman scattering signal due to strong interaction of the medium with pump laser, again caused by the light trapping geometry in addition to the increased total volume of WS$_2$. In addition, we also observe a defect-bound Raman mode (LA(M): 176 cm$^{-1}$) appearing for the N=5 structure of the WS$_2$/h-BN superlattice (Figure 3e). This is likely due to increased defect accumulation as number of layer transfers and thermal cycles of ALD increases. Despite this, the characteristic Raman modes, such as 2LA(M) and A$_{1g}$, maintain constant positions and separations as a function of N, as plotted in Figure 3f.

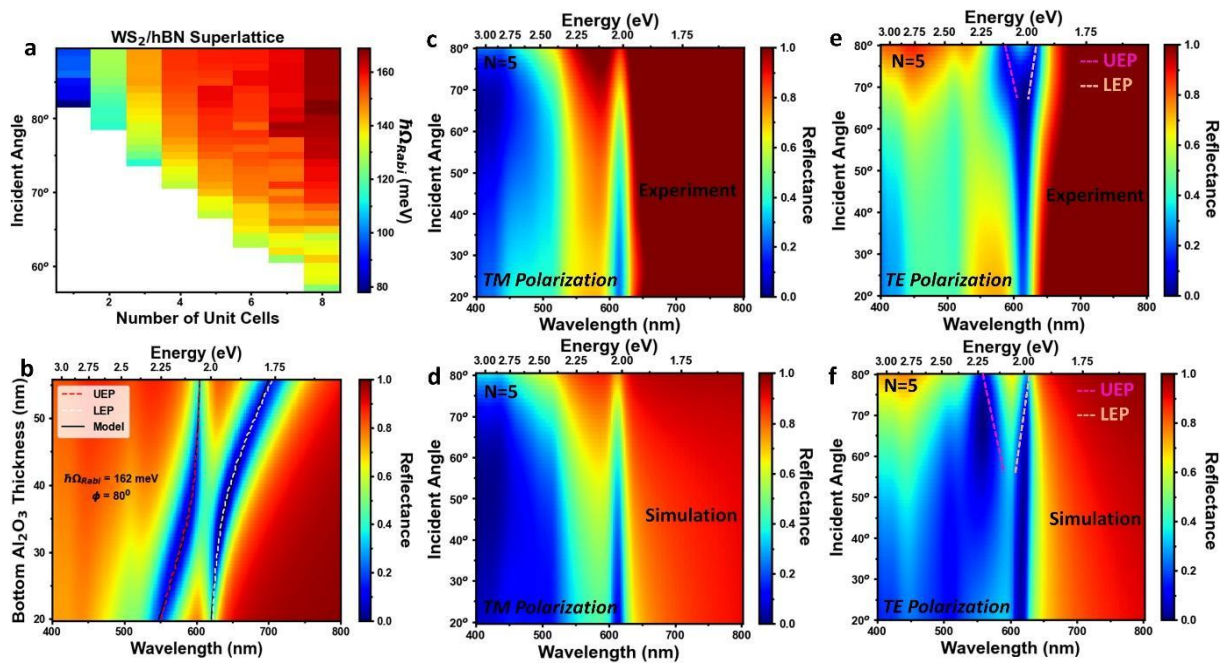

**Figure 4: Observation of exciton-polaritons in un-patterned multilayer superlattices. (a) The simulated Rabi splitting dependence on the incident angle and number of unit cells (b) shows reflectance spectrum dependence on the bottom alumina layer thickness at an incident angle of 80°, which shows the characteristic anti-crossing behaviour of exciton-polaritons. (c-d) The experimental and simulated reflectance spectra, respectively, for various angles of incidence of transverse magnetic (TM) polarized light. (e-f) Experimental and simulated reflectance spectra, respectively, for various angles of incidence of transverse electric (TE) polarized light showing the exciton-polariton splitting starting at an angle of 50°. The upper (UEP) and lower (LEP) exciton-polariton peaks are indicated with dashed lines.**

Thus far, our discussion concerning the light trapping and optical properties of these multilayer superlattices has been focused on normal incidence illumination. We now explore angle dependent coupling of light into this multilayer superlattice structure. Standard TMM simulations show that the excitons in the superlattice hybridize with cavity modes to form exciton-polaritons, with their splitting energy depending on the incident angle and number of unit cells (*Figure 4a*). With a fixed number of unit cells and angle of incidence, a characteristic anti-crossing behavior, signifying exciton coupling with polaritons, can also be seen when the cavity resonance is tuned by varying bottom alumina layer thickness. We use a coupled oscillator model to fit to the simulation data and calculate the Rabi splitting of the system. Our analytical model is in good agreement with the simulated UEP and LEP energies (see supporting info and detailed discussion, Figure S20-24). The Rabi splitting can be increased to 170 meV for a stack of N=8 at >80° incident angles. The Rabi splitting can also be achieved at lower incident angles as the number of unit cells is increased, due to a sharpening in the cavity mode peak, which indicates a lower mode volume, $V_m$ (based on the pathlength of the light inside the superlattice for N layers). The lower mode volume for increased unit cells also results in higher Rabi splitting since the Rabi splitting is inversely proportional to the $\sqrt{V_m}$. The increased Rabi splitting with increased incident angles is due to sharpening in the cavity mode peak, which can be observed in the increased Q-factor (see supporting Info, Figure S23).

This Rabi-splitting is also observed in our superlattice samples and matches well with the simulations (Figure 4c-f). Furthermore, this splitting appears only under TE polarized light (Figure 4 e-f) and not under TM polarized illumination (Figure 4 c-d). This is because monolayer TMDCs can only support TE waveguide modes, unlike their bulk counterparts which can support both TE and TM waveguide modes[18]. As the reflection spectra are acquired away from normal incidence the exciton mode stays unperturbed up until an incident angle of 50° in the simulation. Thereafter, the cavity-induced splitting of the exciton mode emerges with increased incident angles. The emergence of the cavity mode at slightly higher incident angles in experiments when compared to the simulated values can be attributed to presence of polymer and other contamination between the layers, which reduces the quality of the cavity. The Q-factor was obtained by fitting the absorptance peaks to a Lorentzian line shape and was found to follow similar trends to incident angle and number of unit cells as the Rabi splitting (Figure S19). TMM simulations show that the Q-factor of 100 nm thick N=5 $WS_2/Al_2O_3$ superlattices can reach 300 at a moderate incident angle of 70° at room temperature (see supporting Figure S23). The observation of high Q factors in un-patterned multilayer films over square centimetre length scales is potentially valuable for colorimetric sensing applications. In this regime, the exciton-polariton splits into an upper exciton-polariton (UEP) and lower exciton-polariton (LEP) branch for $WS_2$. Both branches were observed simultaneously at room temperature. A third branch was observed in $MoS_2$ since the energy of the A and B excitons are closer than for $WS_2$, and both excitons interacted with the microcavity (see supporting info, Figure S26). The Q-factor of the cavity mode was found to be ~300 at an incident angle of 70° in a subwavelength thick device (100 nm). Since exciton-polaritons maintain light-like and matter-light characteristics, they can be used for non-linear effects in devices such as lasing.

## Conclusions:

In summary, we have demonstrated a centimetre square scale, multilayer superlattice structure based on atomically-thin 2D chalcogenide monolayers acting as quantum wells. The structure of the superlattice was deterministically designed to maximize light trapping at the exciton (~90 %) in < 4 nm thickness of active layer absorber. These superlattices not only maintain a monolayer structure but also support exciton-polaritons at room temperature, with Rabi splitting of up to 170 meV, and cavity modes with quality factors as large as 300 in deep-subwavelength thick devices. Our results show a proof of concept for optical dispersion engineering using atomically thin layers over scalable and arbitrary substrates with broad applications ranging from lasing, sensing as well as optical-modulator devices and lays the foundation towards a materials platform for substrate agnostic integrated nanophotonics.

## Experimental Details:

**Materials and Method:** Uniform and wafer scale TMDCs (monolayer $WS_2$ and $MoS_2$) were grown on c-plane sapphire substrates by MOCVD. Details of the monolayer $WS_2$ and $MoS_2$ used for $WS_2/Al_2O_3$ and $MoS_2/Al_2O_3$ superlattice samples have been published earlier[29,30]. The monolayer $WS_2$ and h-BN (3 nm thick) samples for $WS_2$/h-BN superlattice were grown by Aixtron Ltd. (UK) using a Close Coupled Showerhead® metal organic vapour deposition reactor using tungsten hexacarbonyl and ditertiarybutylsulfide, and borazine respectively. Atomic layer deposition (ALD) of $Al_2O_3$ layer was performed with Cambridge Nanotech (USA) where metal organic precursor of TMA was used with water vapor in each cycle. PVD of Au/Ti (100/10 nm) films were done with the e-beam evaporation technique utilizing an instrument manufactured by K. J. Lasker, USA. Mechanically exfoliated samples were prepared using the scotch tape method and placed onto reflective substrates with the help of PDMS stamp using a dry transfer technique. Bulk $WS_2$ crystal purchased from hq-Graphene (Netherlands) was used for the exfoliation of few layered sample. Wafer-scale few-layer $MoSe_2$ thin films were synthesized by first sputtering of Molybdenum (Mo) via an asymmetric bi-polar pulsed direct current magnetron sputtering system at 65 KHz (0.4 sec reverse time) from a Mo target. A sputtering time of 4 seconds resulted in a uniform atomically smooth Mo film with thicknesses of 0.6 nm. Following the Mo deposition, the thin metal films were then transferred into a hot wall CVD reactor evacuated and purged with a flow of $H_2$. The films were heated to 650° C under a flow of $N_2/H_2$ (95%:5%). After reaching 650° C, films were selenized under a flow of $H_2Se$ for 30 min then cooled to 400° C before turning off the flow of $H_2Se$ and cooled to room temperature under $N_2/H_2$.

**Characterization:** The surface roughness of the $Au/Al_2O_3$ layer on the $Si/SiO_2$ substrates were analyzed with atomic force microscopy (AFM; AIST, USA). As received MOCVD grown monolayer TMDCs as well as 3 nm thick h-BN samples were analyzed with Raman, photoluminescence (PL) and Reflectance spectroscopy, performed at integrated system available in Horiba Scientific Confocal Microscope (LabRAM HR Evolution). This instrument is equipped with an Olympus objective lens (up to 100x) and three different grating (100, 600 and 1800) based spectrometers, which are coupled to a Si focal plane array (FPA) detector. A continuous-wave excitation source with excitation wavelength at 405 nm and 633 nm was used to perform PL and Raman spectroscopy studies, respectively. 10% laser power of 405 nm wavelength corresponds to 22 micro Watt while utilizing 100x objective lens for power dependent study. Visible white light incident through the fibre probe was utilized for reflectance spectral analysis, using the Horiba Scientific confocal microscope. Temperature-dependent spectroscopic analysis was performed on the same Horiba instrument utilizing Linkam heating/cooling stage where temperature was precisely calibrated by the equilibration time. Spectral spectroscopic ellipsometry analysis of all the samples was performed using a J. A. Woollam Ellipsometer (Model: M-2000 – detector spectral range of 371–1687 nm) to obtain optical constants. Similarly, multi-incidence, angle-dependent reflectance spectral behaviour was also analyzed with Spectroscopic Ellipsometry (SE). ABF-STEM, TEM and EDS measurements were performed at 200 keV using Transmission Electron

Microscopes (F200-JEOL). Images were captured by a Gatan annular detector using Gatan's GMS Software. Cross-sectional samples were prepared with Xe$^+$ plasma based focused ion beam / scanning electron microscope (S8000X-TESCAN).

**Theoretical Modelling:**

A 2x2 transfer matrix method[31] was used to simulate the reflectance of the superlattice structure/heterostructures. We adopted the open-source TMM code[32] and further developed it in MATLAB$^{TM}$. The TMM was used to simulate the Rabi splitting, and a coupled oscillator model fit was performed to calculate the Rabi energy. The coupled oscillator models the system using a 2x2 Hamiltonian where the diagonal terms are the undisturbed microcavity and exciton energies while the off-diagonal terms are the Rabi energy and characterize the strength of the interaction between the excitonic and optical states. A three-body coupled oscillator model was used for $MoS_2$ and $MoSe_2$ (see supporting Info, Figure S22 and S27).

## Author contributions:

D.J. and A.R.D. conceived the idea/concept. D.J. directed the collaboration and execution. F.B. and J.L. optimized the superlattice design computationally. P.K. reduced the design to experiments including performing all superlattice fabrication, mechanical exfoliated sample preparation, Cross-sectional TEM sample preparation, optical spectroscopy (including low temperature) and electron microscopy characterization. J.L. with help of H.L. and J.D. performed simulations and fitting the experimental data to computational models. B.S. assisted with measurement, optimization, and fitting of optical constants. H. Zhang and S.B.A. assisted in sample preparation and characterization respectively. M.K. assisted with theoretical modelling and interpretation. H. Zhu, T.H.C. and J.R. synthesized $MoS_2$ and $WS_2$ samples used in $Al_2O_3$ spaced superlattices. C.McA., X.W., B.R.C and O.W. led the synthesis of $WS_2$ and h-BN used in the h-BN spaced superlattice samples. M.J.M., M.S., C.M. and N.J.G. synthesized and characterized the $MoSe_2$ samples. E.A.S. supervised the electron microscopy experiments. P.K., J.L. and D.J. wrote the manuscript and all authors contributed to the writing and editing of the manuscript.

## Acknowledgements:

D.J. acknowledges primary support for this work by the U.S. Army Research Office under contract number W911NF-19-1-0109 and Air Force Office of Scientific Research (AFOSR)

FA9550-21-1-0035. D.J. and J.L. also acknowledge partial support from and FA2386-20-1-4074 and the University Research Foundation (URF) at Penn. D.J., E.A.S. and P. K. acknowledge support from National Science Foundation (DMR-1905853) and support from University of Pennsylvania Materials Research Science and Engineering Center (MRSEC) (DMR-1720530) in addition to usage of MRSEC supported facilities. The sample fabrication, assembly and characterization was carried out at the Singh Center for Nanotechnology at the University of Pennsylvania which is supported by the National Science Foundation (NSF) National Nanotechnology Coordinated Infrastructure Program grant NNCI-1542153. F.B. is supported by the Vagelos Integrated Program in Energy Research. H.Z. was supported by Vagelos Institute of Energy Science and Technology graduate fellowship. S.B.A acknowledges support from Swiss National Science Foundation Early Postdoc Mobility Program (187977). A.R.D. acknowledges support from Northrop Grumman and UCLA Council on Research Faculty Research Grant. The TMDC monolayer samples were provided by the 2D Crystal Consortium–Materials Innovation Platform (2DCC-MIP) facility at the Pennsylvania State University which is funded by NSF under cooperative agreement DMR-1539916. M.S. and N.R.G. acknowledge support from the Air Force Office of Scientific Research under Award No. FA9550-19RYCOR050. The Author's acknowledge helpful discussion on light-coupling with Mark W. Knight.

# Supporting Information

## Light-Matter Coupling in Scalable Van der Waals Superlattices

Pawan Kumar[1,2], Jason Lynch[1], Baokun Song[1$], Haonan Ling[3], Francisco Barrera[1], Huiqin Zhang[1], Surendra B. Anantharaman[1], Jagrit Digani[3], Haoyue Zhu[4], Tanushree H. Choudhury[4], Clifford McAleese[5], Xiaochen Wang[5], Ben R. Conran[5], Oliver Whear,[5] Michael J. Motala[6], Michael Snure[7], Christopher Muratore[8], Joan M. Redwing[4], Nicholas R. Glavin[6], Eric A. Stach[2], Artur R. Davoyan[3], Deep Jariwala[1*]

[1]Electrical and Systems Engineering, University of Pennsylvania, Philadelphia, PA, 19104, USA
[2]Materials Science and Engineering, University of Pennsylvania, Philadelphia, PA, 19104, USA
[3]Department of Mechanical and Aerospace Engineering, University of California at Los Angeles, CA, 90095, USA
[4]2D Crystal Consortium-Materials Innovation Platform, Materials Research Institute, Pennsylvania State University, University Park, PA, 16802, USA
[5]AIXTRON Ltd., Anderson Road, Swavesey, Cambridge, CB24 4FQ, UK
[6]Air Force Research Laboratory, Materials and Manufacturing Directorate, Wright-Patterson AFB, Dayton, OH, 45433, USA
[7]Air Force Research Laboratory, Sensors Directorate, Wright-Patterson AFB, Ohio 45433, USA
[8]Department of Chemical and Materials Engineering, University of Dayton, Dayton, Ohio, 45469, USA

*Corresponding author: dmj@seas.upenn.edu
$Present Address: Huazhong University of Science and Technology, China

## Superlattice Fabrication Details:

**Wet-chemical transfer and stacking of van der Waals TMDCs layers:** The wet chemical transfer technique (see supporting info Figure S2) was utilized throughout the stacking process of large area (~1x1 cm$^2$) atomically thin TMDCs layers grown under MOCVD. As received wafer scale uniformly grown 2-dimensional monolayer of MoS$_2$, WS$_2$ and 3 nm thin h-BN on 2" sapphire (single side polished) was first cut into several pieces of dimension 1x1 cm$^2$. Poly Methyl Metha Acrylate (PMMA) 950k A4 was used to spin coat on these 1x1 cm$^2$ based TMDCs samples and kept in air to dry overnight to prepare for wet chemical delamination process of atomic thin layers. PMMA thin film was deposited with a controlled thickness of ~200 nm. PMMA coated samples were dipped in de-ionized (DI) water heated to 85° C on a hot plate, for 20-30 minutes until air-bubbles began to form at the outer edges of the sapphire substrate. Following the formation of air bubbles, samples were taken out of the hot water and placed on the top of 3M KOH solution which was maintained at 85° C. Crucial delamination of PMMA-supported TMDC layers from sapphire substrate occurred at this step where sapphire substrate was held manually with 45° inclination and slowly dipped inside the KOH solution with extremely slow movement. This allowed the delamination of the PMMA-supported TMDCs layers which finally floated to the top of the KOH solution. Using a cleaned glass slide, the floating PMMA-supported TMDCs layers were transferred to the fresh DI water, and this step was repeated multiple times to remove any residual contamination from the delamination steps. Finally, floating PMMA supported TMDCs were scooped on desired Si substrate covered with Au (Si/SiO$_2$/Au/Al$_2$O$_3$) and left to air-dry overnight. The final lift-off process (to remove the PMMA film) was done with the help of acetone for 6 hours at moderate temperature (45° C). To allow for better adhesion of the wet-chemical transferred TMDCs layers to their new substrate, they were heated for 1-2 hours at 70° C on a hot plate. The processes described above for producing the multilayer stacks including precise manual alignment at each scooping step were followed for the synthesisation of each heterostructure. Alumina-based superlattices were fabricated with additional steps of depositing a Al$_2$O$_3$ layer by ALD after each wet-chemical transfer of the TMDC layer. The topmost TMDCs layer in all superlattices were not coated with either Al$_2$O$_3$ or h-BN.

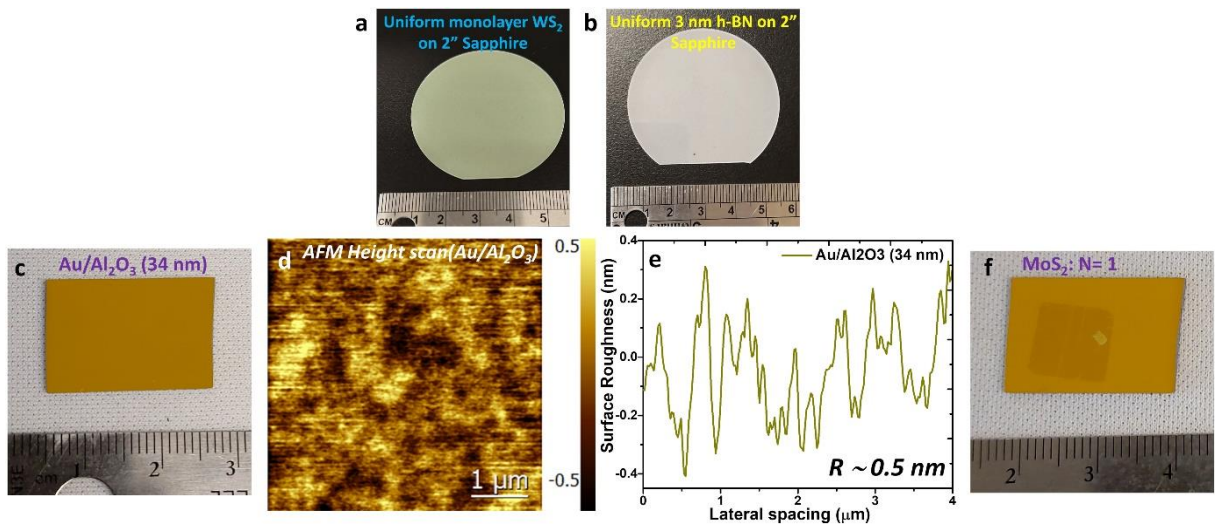

**Figure S1:** Camera clicked optical images of the 2" wafers of (a) monolayer WS$_2$ and (b) 3 nm thick h-BN. (c) Picture of 34 nm of atomic layer deposited Al$_2$O$_3$ on top of e-beam evaporated Au. (d) The atomic force microscopic (AFM) height image of the Al$_2$O$_3$/Au substrate and corresponding (e) Surface profile with measured surface roughness of ∼0.5 nm. (e) Picture of monolayer MoS$_2$ transferred on to Al$_2$O$_3$/Au surface.

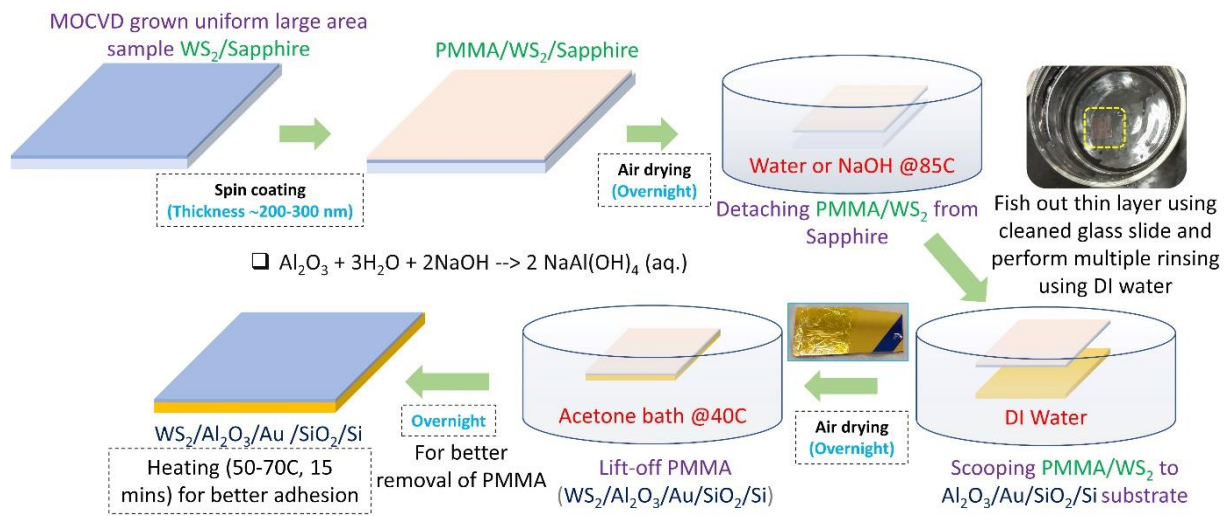

**Figure S2:** Wet chemical transfer technique shown with sequential steps needed with the appropriate time accounted for best result. Monolayer WS$_2$ as grown on c-plane sapphire substrate by MOCVD has been shown to transfer to new substrate (Al$_2$O$_3$/Au/SiO$_2$/Si).

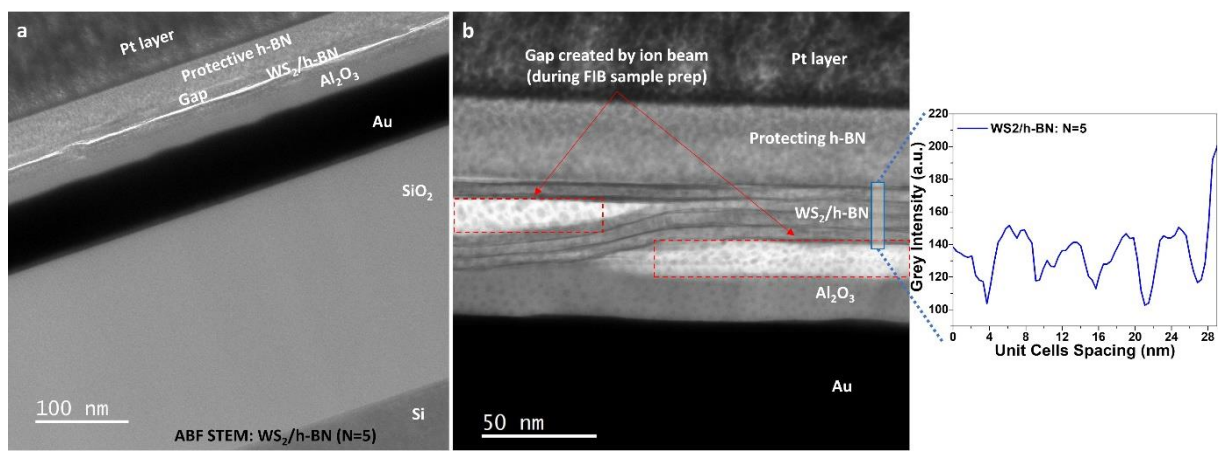

**Figure S3:** (a) Low magnification annular bright field (ABF) scanning transmission electron microscope (STEM) imaging of the WS$_2$/h-BN superlattice (N=5) covering the entire cross-sectional view. The protective h-BN and Pt layers were deposited on top to provide suitability of the plasma-FIB lamella preparation. (b) Magnified ABF STEM image shows the superlattice region where white contrast appears in the mid-section of the image formed due to air gap created by FIB milling process and corresponding line profile across the WS$_2$/h-BN unit cells as marked through blue rectangular region.

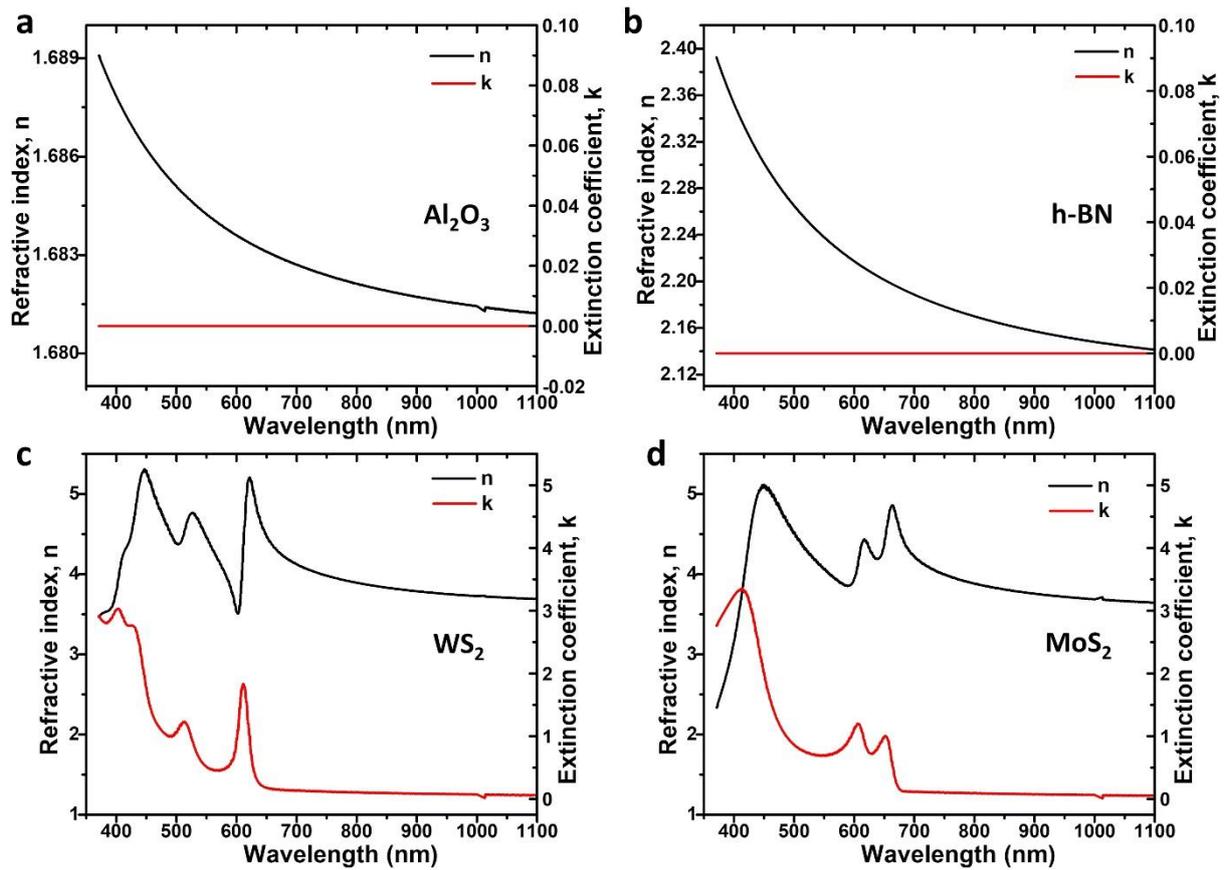

**Figure S4:** The measured complex refractive index of (a) $Al_2O_3$, (b) h-BN, (c) $WS_2$, and (d) $MoS_2$. All were obtained using angle-resolved ellipsometry on single side polished sapphire. Samples that were grown on double side polished substrates were transferred on single side polished sapphire substrates before ellipsometry measurements and analysis. $Al_2O_3$ and h-BN were fitted using a Cauchy model, and $WS_2$ and $MoS_2$ were fitted using a series of Tauc-Lorentz oscillators. All optical constants were measured on single side polished sapphire substrates.

## TMM Modeling and Genetic Algorithm Optimization:

Initial Transfer Matrix Method (TMM) modelling was carried out using a MATLAB script provided by the McGehee group[1] adapted to incorporate the repeating unit cell scheme. The standard script allows for the simulation of a multilayer superlattice in which the material, thickness, and relative order of each layer may be specified. Additionally, the script reads real and complex refractive index values for each properly referenced material from an external excel file. The script was converted to python and modified to allow for quicker simulations and angled incidence of both TE and TM polarized light. The algorithm follows the approach described by Petterson *et al*[2].

A genetic-algorithm-based optimization was implemented with the figure of merit being the primary excitonic absorptance for $WS_2$-based superlattices and the average absorptance of the primary and secondary excitons for $MoS_2$ and $MoSe_2$-based superlattices. Holding all other inputs constant, inputs controlling the number of unit cells, the thickness of the unit cell spacers, and the thickness of the bottom spacer were allowed to vary in this optimization.

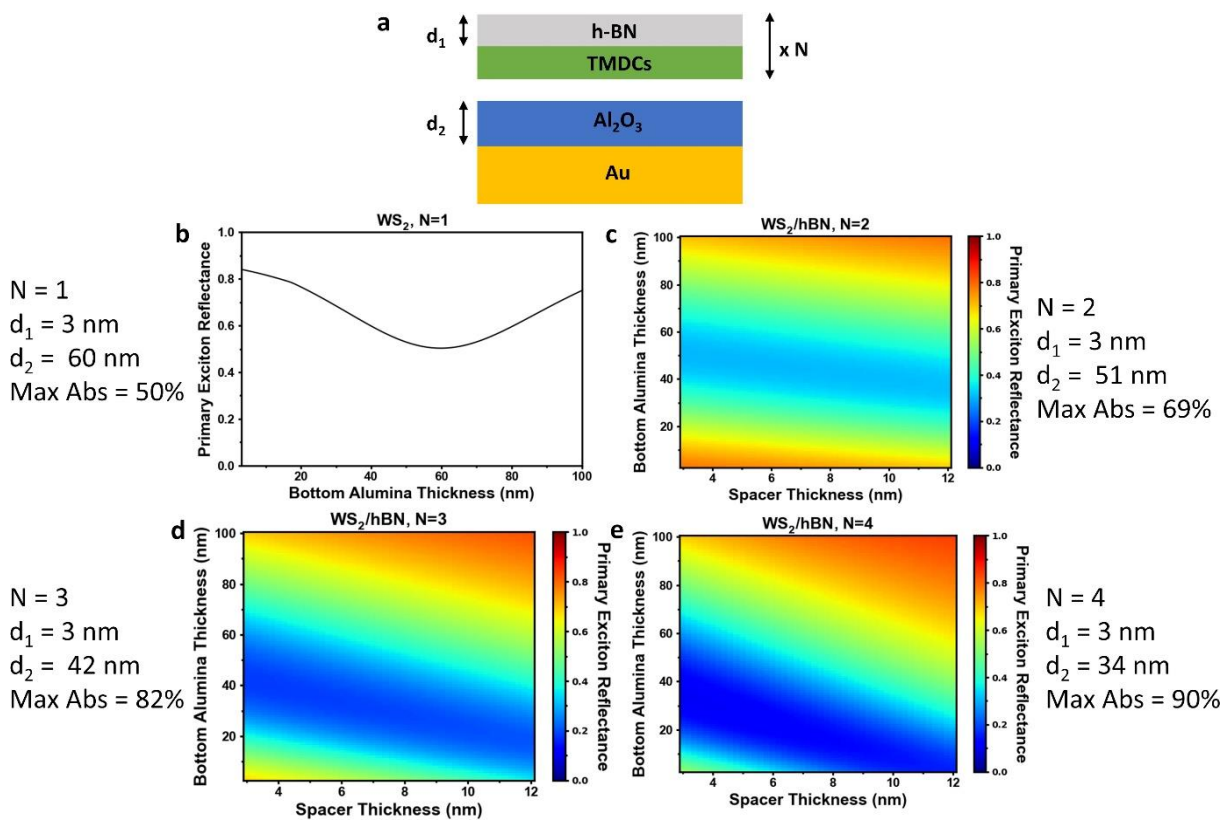

**Figure S5:** The results for optimizing the $WS_2/Al_2O_3$ superlattice for N=1 through N=4 for maximum primary exciton (613 nm) absorptance which corresponds to a minimum in the reflectance. (a) Schematic of our device with the optimized parameters marked. $d_1$ is called the spacer thickness, and $d_2$ is the bottom alumina thickness. (b-e) The optimization results for varying spacer and bottom alumina thickness with the optimal parameters listed for N=1 through 4.

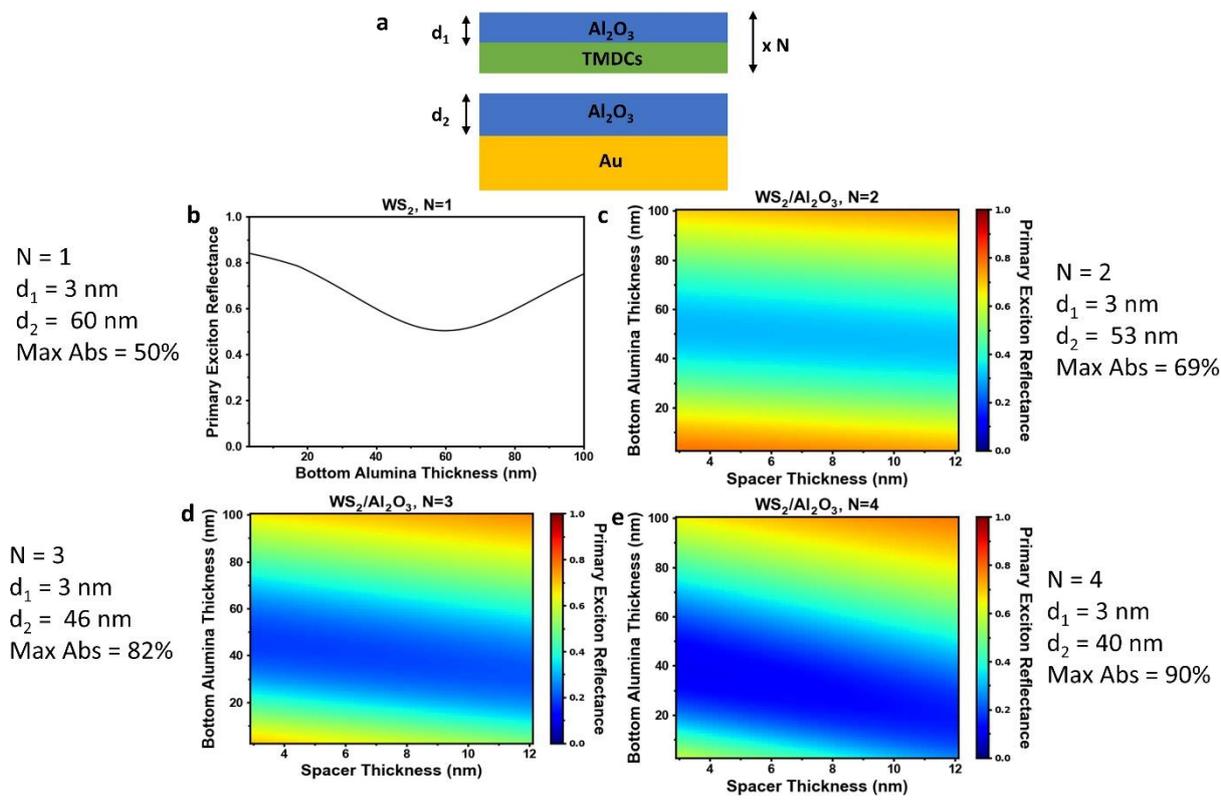

**Figure S6: The results for optimizing the MoS$_2$/Al$_2$O$_3$ superlattice for N=1 through N=4 for maximum average absorptance of the A and B excitons (575 to 675 nm). (a-d) The optimization results for varying spacer and bottom alumina thickness with the optimal parameters listed for N=1 through 4. d$_1$ is the spacer thickness, and d$_2$ is the bottom alumina thickness.**

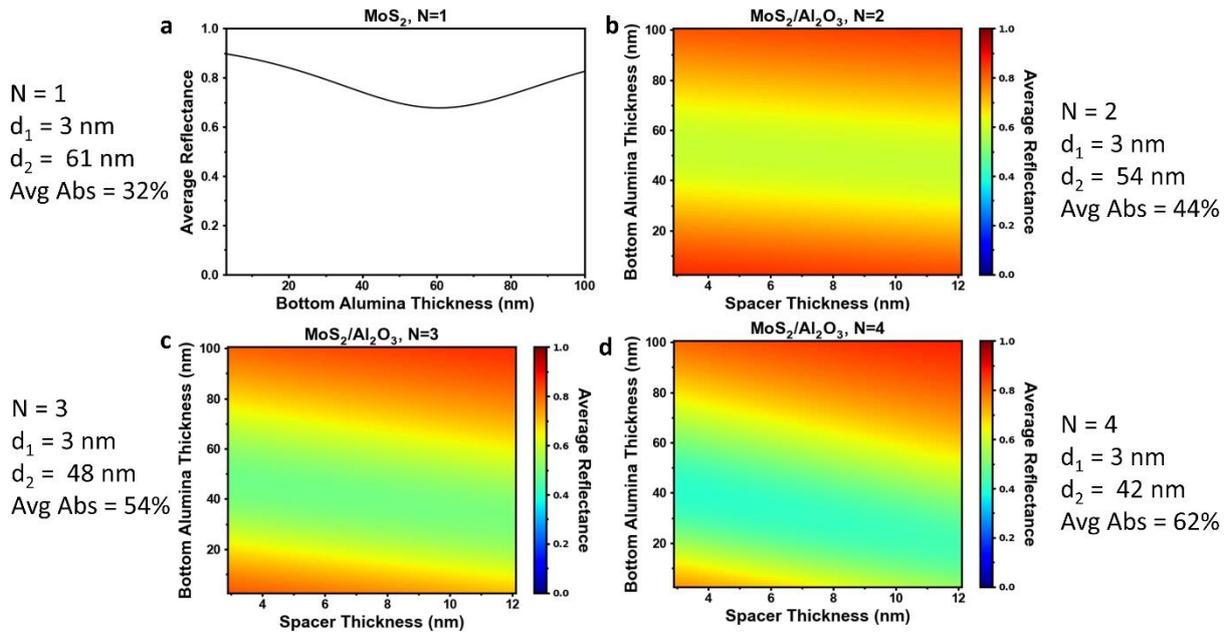

Figure S7: The results for optimizing the WS$_2$/hBN superlattice for N=1 through N=4 for maximum primary exciton (613 nm) absorptance which corresponds to a minimum in the reflectance. (a) Schematic of our device with the optimized parameters marked. d$_1$ is called the spacer thickness, and d$_2$ is the bottom alumina thickness. (b-e) The optimization results for varying spacer and bottom alumina thickness with the optimal parameters listed for N=1 through 4.

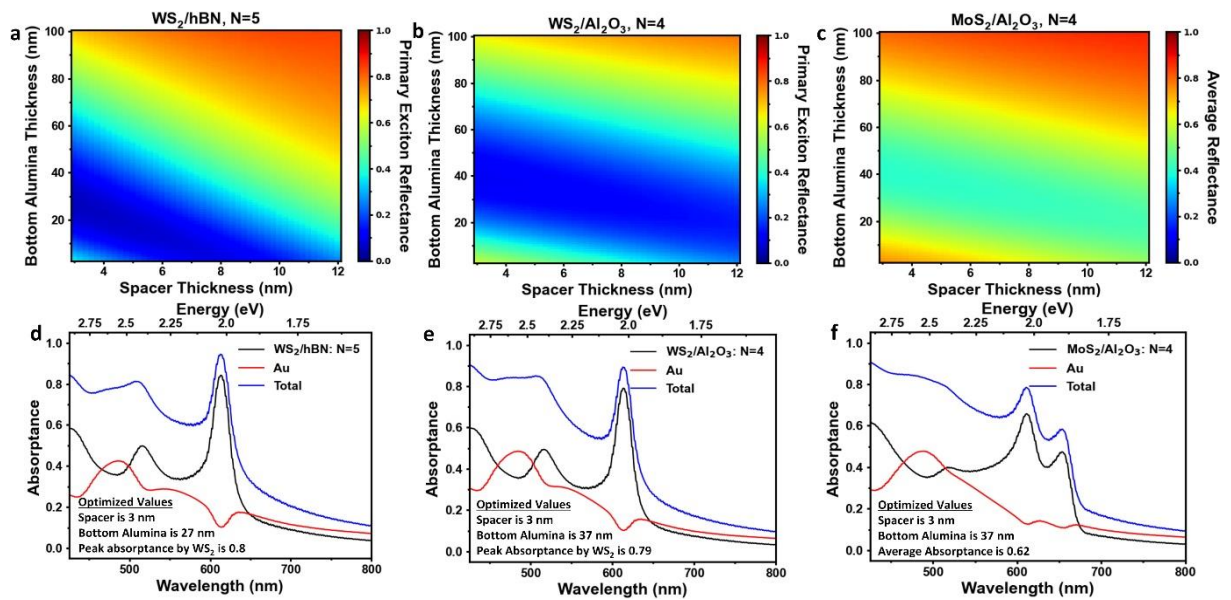

Figure S8: The optimization for the fabricated superlattices of (a) WS$_2$/h-BN (N=5), (b) WS$_2$/Al$_2$O$_3$ (N=4), and (c) MoS$_2$/Al$_2$O$_3$ (N=4) for varying bottom alumina and spacer thicknesses. The layer resolved absorptances of the optimized superlattices are shown in (d-f) for WS$_2$/h-BN (N=5), WS$_2$/Al$_2$O$_3$ (N=4), MoS$_2$/Al$_2$O$_3$ (N=4), respectively. The optimized parameters and absorptance values are indicated in the graphs.

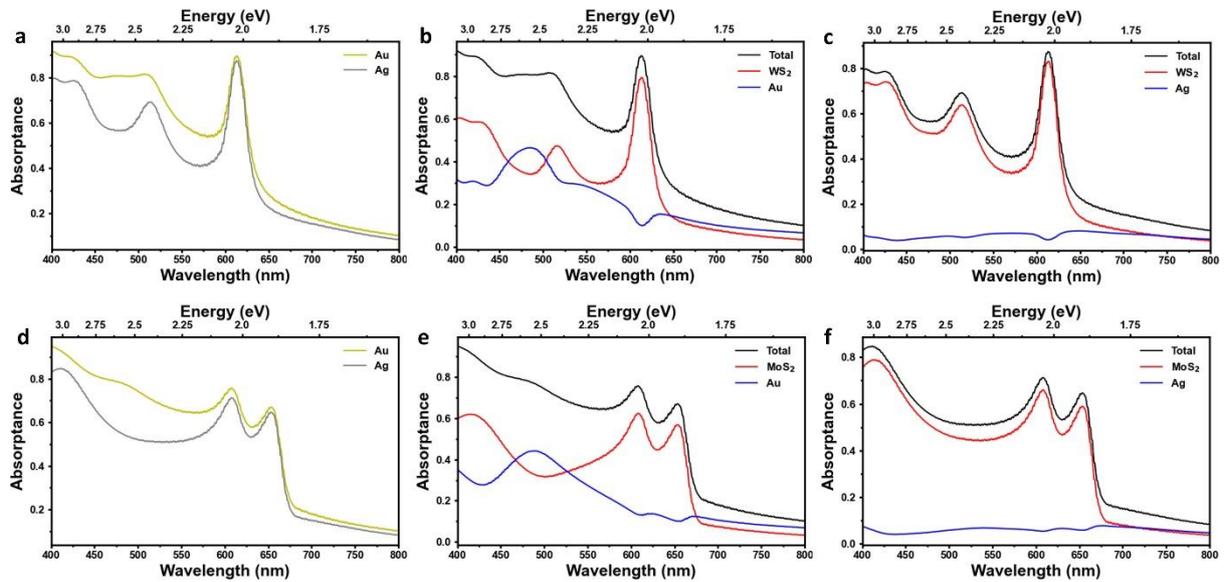

**Figure S9:** The absorptance dependence for superlattices with Au and Ag reflectors. (a, d) Comparison of the optimized spectra for $WS_2/Al_2O_3$: N=4 and $MoS_2/Al_2O_3$: N=4 superlattices, respectively. (b, c) The layer resolved absorptance for a $WS_2/Al_2O_3$: N=4 superlattice with Au and Ag reflectors. The Au has an absorptance of 0.10 at the primary exciton wavelength (613 nm) while the Ag reflector has an absorptance of 0.04 at the same wavelength. (e, f) The layer resolved absorptance for a $MoS_2/Al_2O_3$: N=4 superlattice with Au and Ag reflectors. The Au reflector has absorptances of 0.1 and 0.14 at the A (655 nm) and B (608 nm) exciton wavelengths, respectively, while the Ag reflector has absorptances of 0.06 and 0.06 as the same wavelengths.

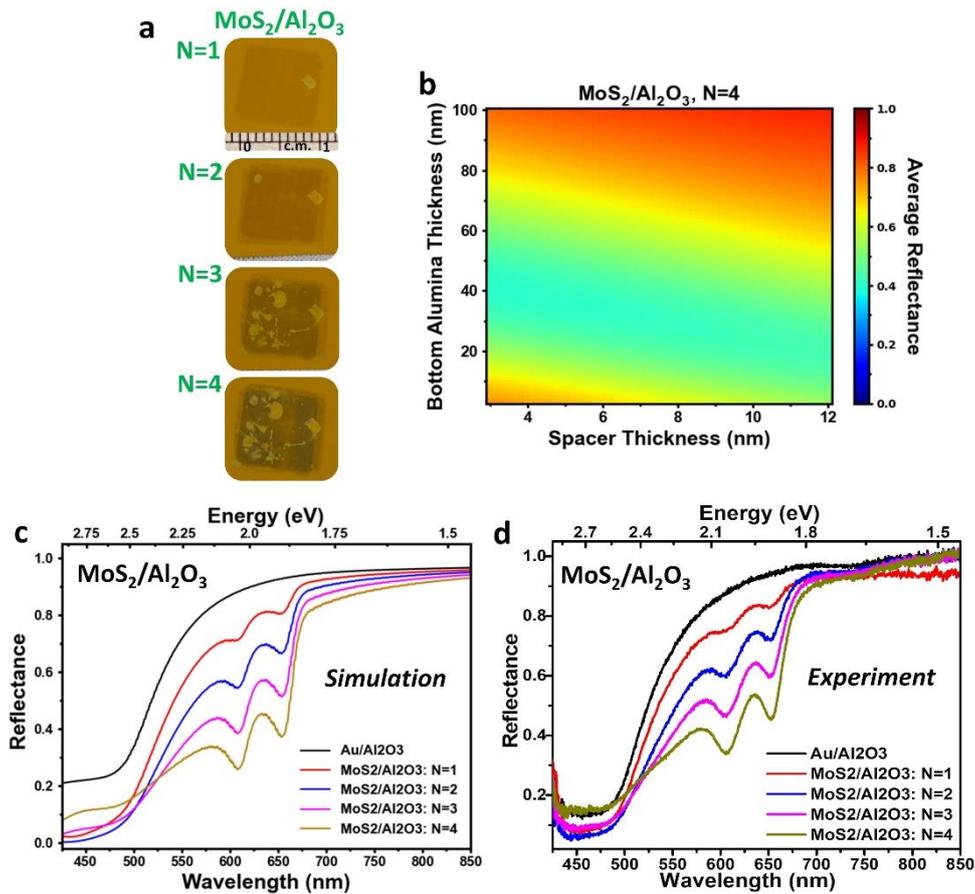

**Figure S10:** (a) Show the deposition of different number of unit cells of the $MoS_2/Al_2O_3$ superlattice. (b) The optimization process for the $MoS_2/Al_2O_3$ N=4 superlattice. (c) The simulated reflectance of the $Al_2O_3$/Au substrate and varying number of unit cells, and (d) the experimentally measured reflectance of the $Al_2O_3$/Au substrate and varying number of unit cells.

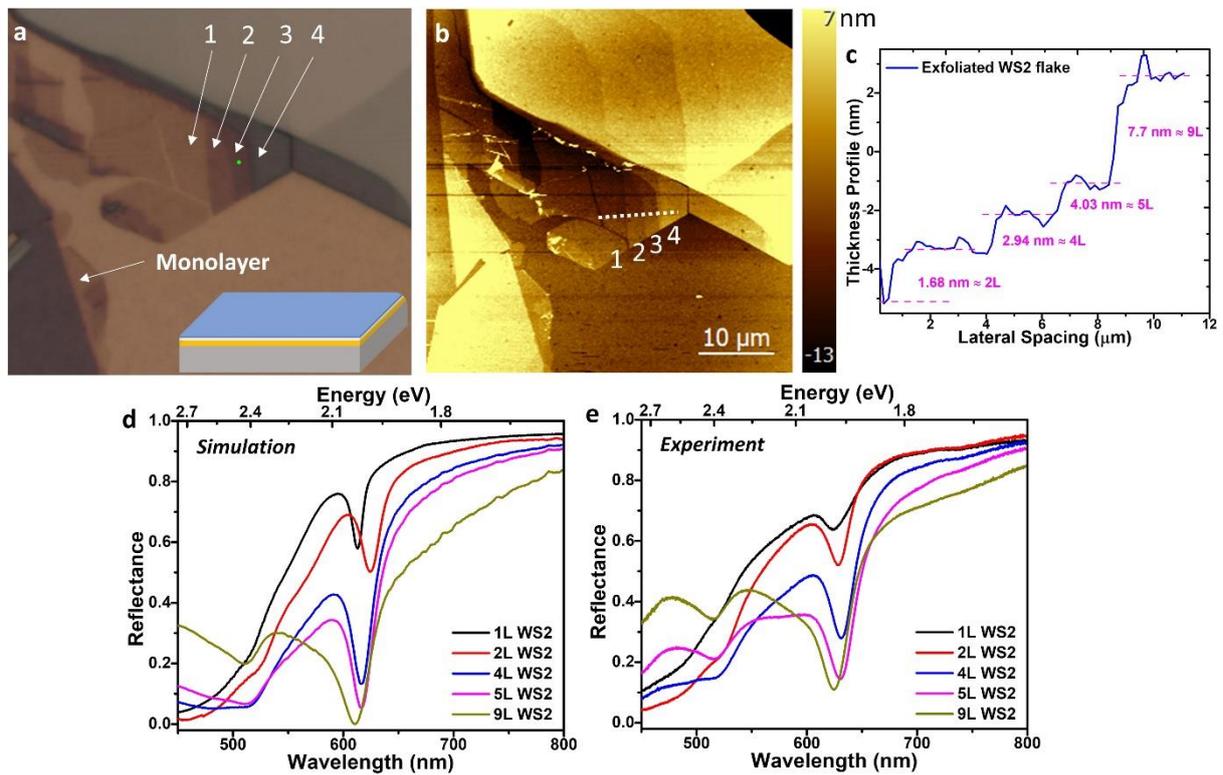

**Figure S11:** (a) A microscopic image of the exfoliated WS$_2$ sample on Al$_2$O$_3$ (34 nm)/Au with the monolayer, bi-layer, 4-layer, 5-layer, and 9-layer regions labelled monolayer, 1, 2, 3, and 4, respectively. (b) AFM image of the same area shown in (a). (c) shows the thickness profile along the dashed line in (b). (d) is the simulated reflectance of the different regions using WS$_2$ complex refractive indices obtained from literature. (d) the experimental reflectance of the regions.

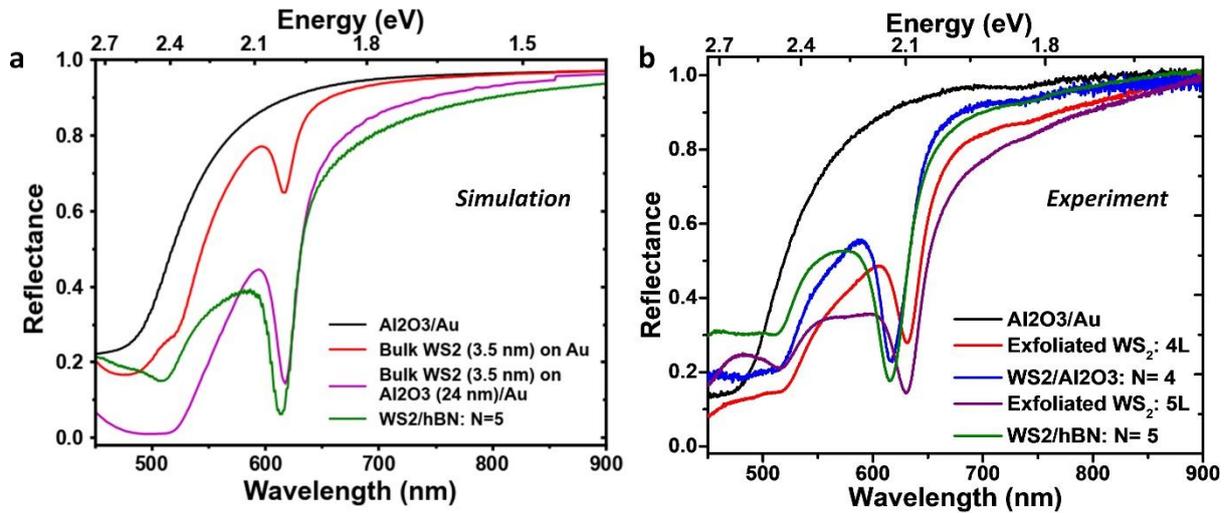

**Figure S12:** (a) The simulated reflectance of the WS$_2$/h-BN N=5 superlattice compared to architectures with similar thicknesses of bulk WS$_2$. (b) The experimental reflectance of the N=4 and N=5 superlattice WS$_2$/h-BN superlattices compared to the exfoliated sample of similar WS$_2$ thicknesses. Blue shift of superlattice peak showing the monolayer confined structure as opposed to bulk exfoliated WS$_2$. For 4L it is pretty clear that monolayer superlattice is superior compared to 4L thick WS$_2$, even for 5L case, superlattice is superior to exfoliated WS$_2$ as is evident from the simulations.

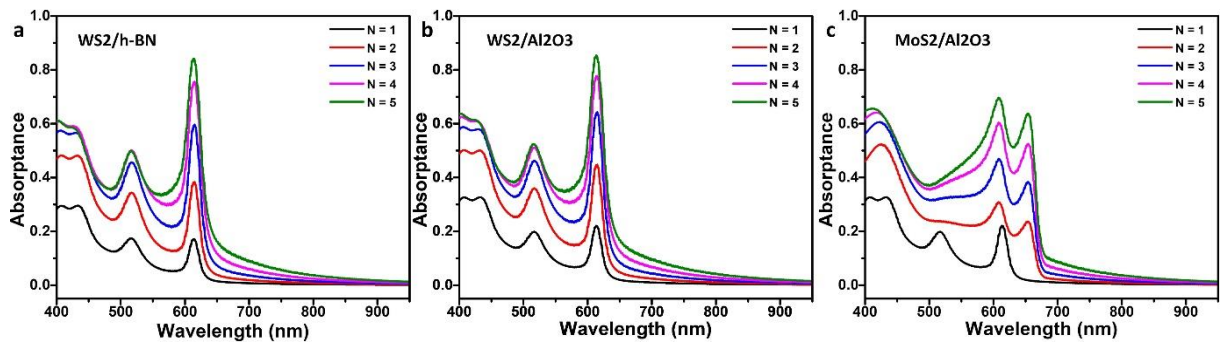

**Figure S13:** (a-c) Layer resolved absolute absorptance of the TMDC layers (WS$_2$ and MoS$_2$) in three different superlattice structure, simulated using TMM model for varying number of unit cells from N=1 to 5.

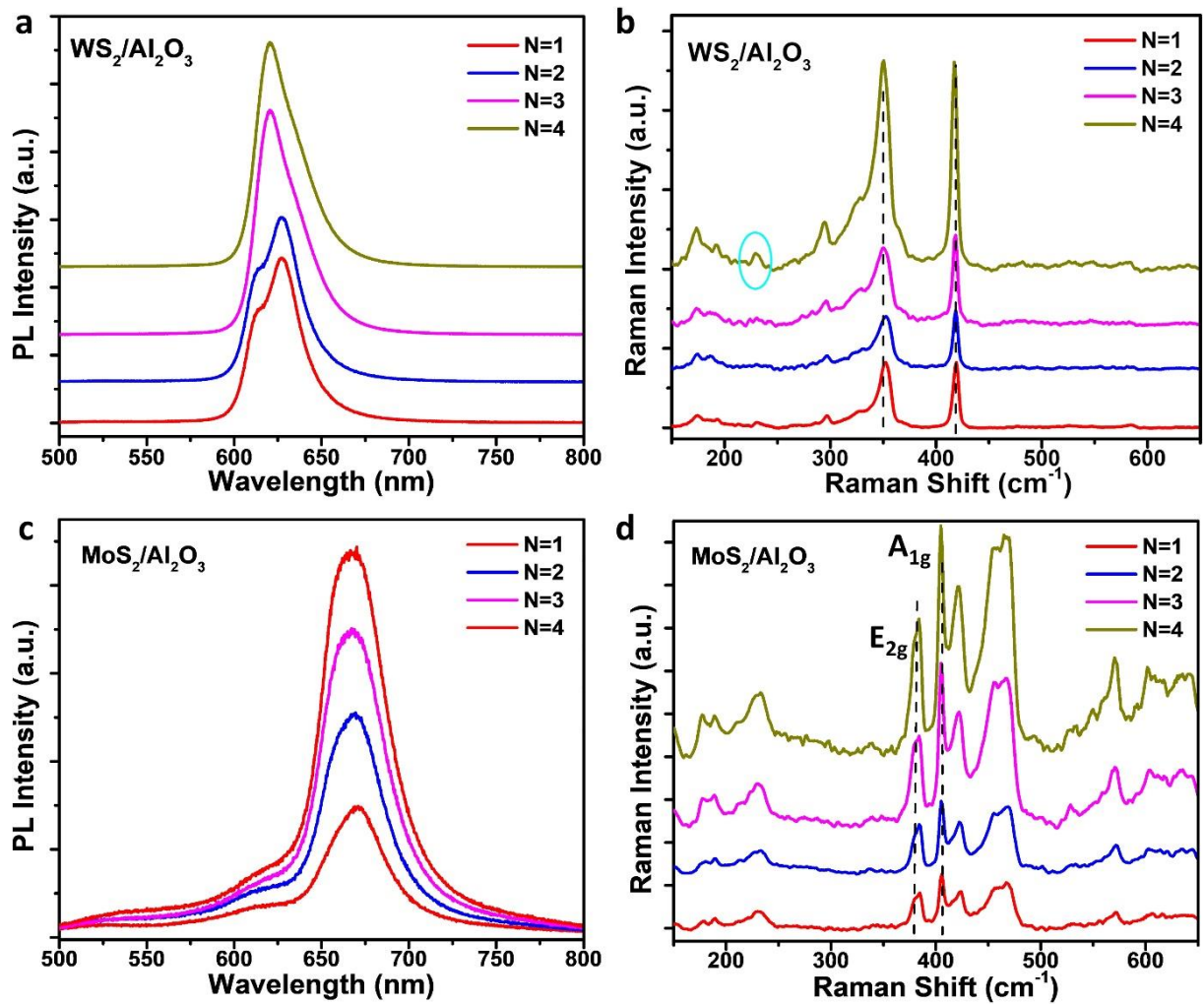

Figure S14: (a) The number of unit cells dependent PL of the $WS_2/Al_2O_3$ superlattice. The exciton and trion peaks merge with increasing N, and the main peak blue shifts. (b) The number of unit cells dependent Raman spectra of the $WS_2/Al_2O_3$ superlattice. The monolayer Raman spectra characteristics are maintained throughout the stacking process. The blue circle shows the emergence of a defect state with increased stacking. (c) The number of unit cells dependent PL of the $MoS_2/Al_2O_3$ superlattice. (d) The number of unit cells dependent Raman spectra of the $MoS_2/Al_2O_3$ superlattice. The monolayer Raman spectra characteristics are maintained throughout the stacking process.

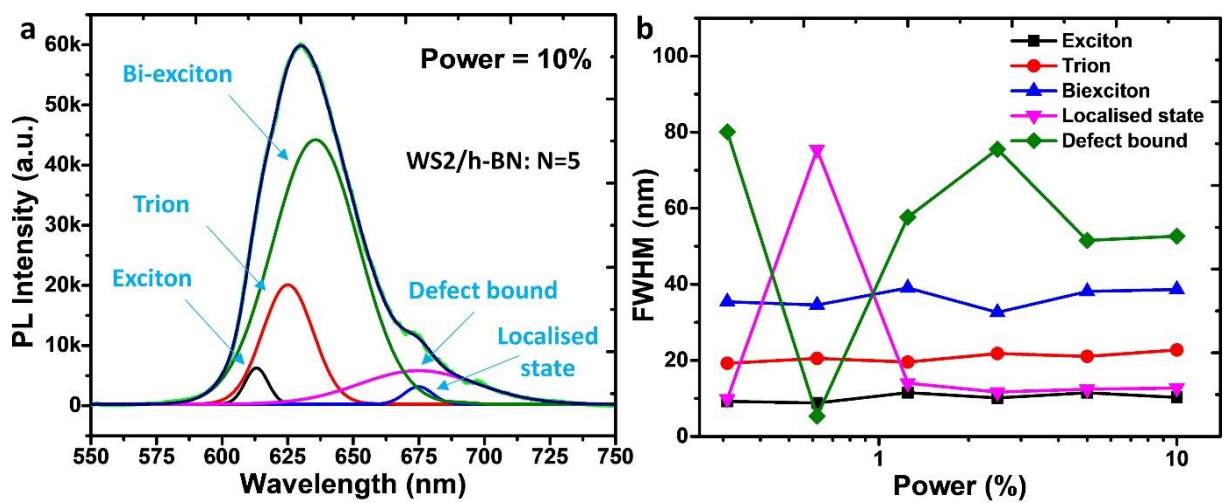

**Figure S15:** (a)The highest power (10%) PL spectrum for WS$_2$/h-BN decomposed into its constituent peaks using a series of Lorentzian multipeak fittings. (b) The line-width dependence on varying power density for the WS$_2$/h-BN: N=5 superlattice.

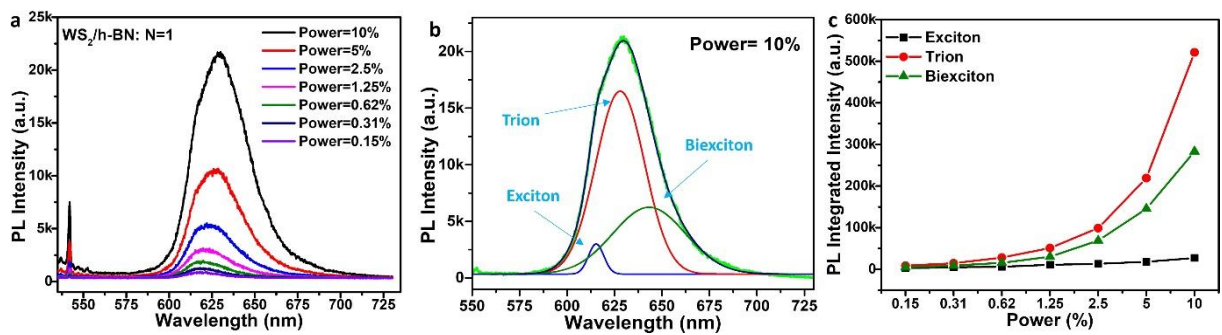

**Figure S16:** (a) The power dependent PL spectra for the WS$_2$/h-BN N=1 superlattice. (b) The highest power PL spectrum for WS$_2$/h-BN decomposed into its constituent peaks using a series of Lorentzian fits. (c) The intensities of the constituent PL excitonic peaks and their dependence on laser power.

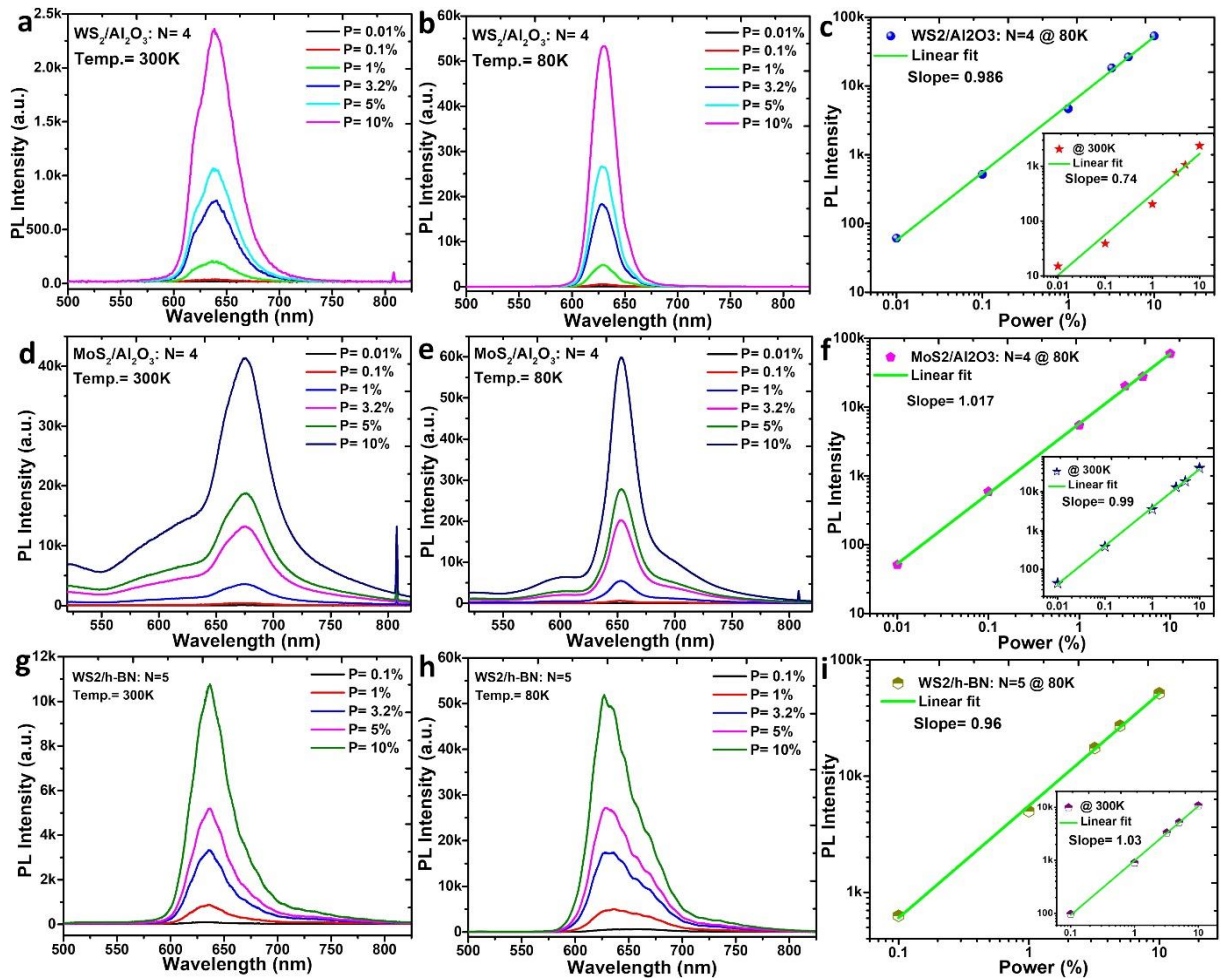

**Figure S17:** The dependent PL spectra for WS$_2$/Al$_2$O$_3$ superlattice at temperatures of (a) 300 K and (b) 80 K. The PL intensity increased by a factor of ~23 when cooling the sample from 300 K to 80 K. (c) Shows the linear dependent of the maximum PL intensity on laser power for the WS$_2$/Al$_2$O$_3$ N=4 superlattice at 80 K. The dependent PL spectra for MoS$_2$/Al$_2$O$_3$ superlattice at temperatures of (d) 300 K and (e) 80 K. (f) Shows the linear dependent of the maximum PL intensity on laser power for the MoS$_2$/Al$_2$O$_3$ N=4 superlattice at 80 K. The dependent PL spectra for WS$_2$/h-BN superlattice at temperatures of (g) 300 K and (h) 80 K. The PL intensity increased by a factor of ~5 when cooling the sample from 300 K to 80 K. (i) shows the linear dependent of the maximum PL intensity on laser power for WS$_2$/h-BN superlattice. Inset of figure c, f, i consist linear dependencies of laser power upon the PL peak intensities measured at room temperature.

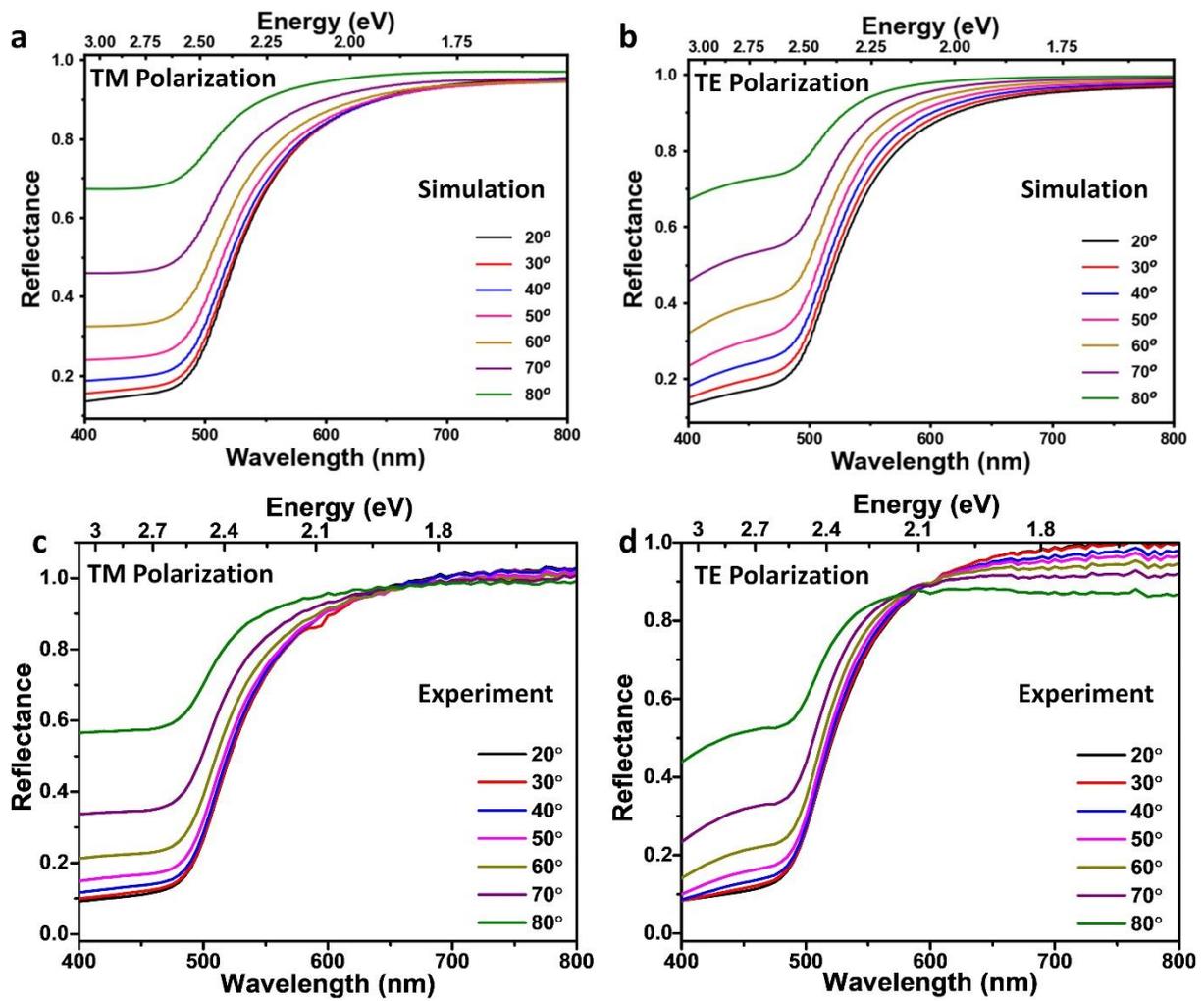

Figure S18: The simulated angle-resolved reflectance of 34 nm of $Al_2O_3$ on Au for (a) TM and (b) TE polarized light. The corresponding experimental spectra for (c) TM and (d) TE polarized light.

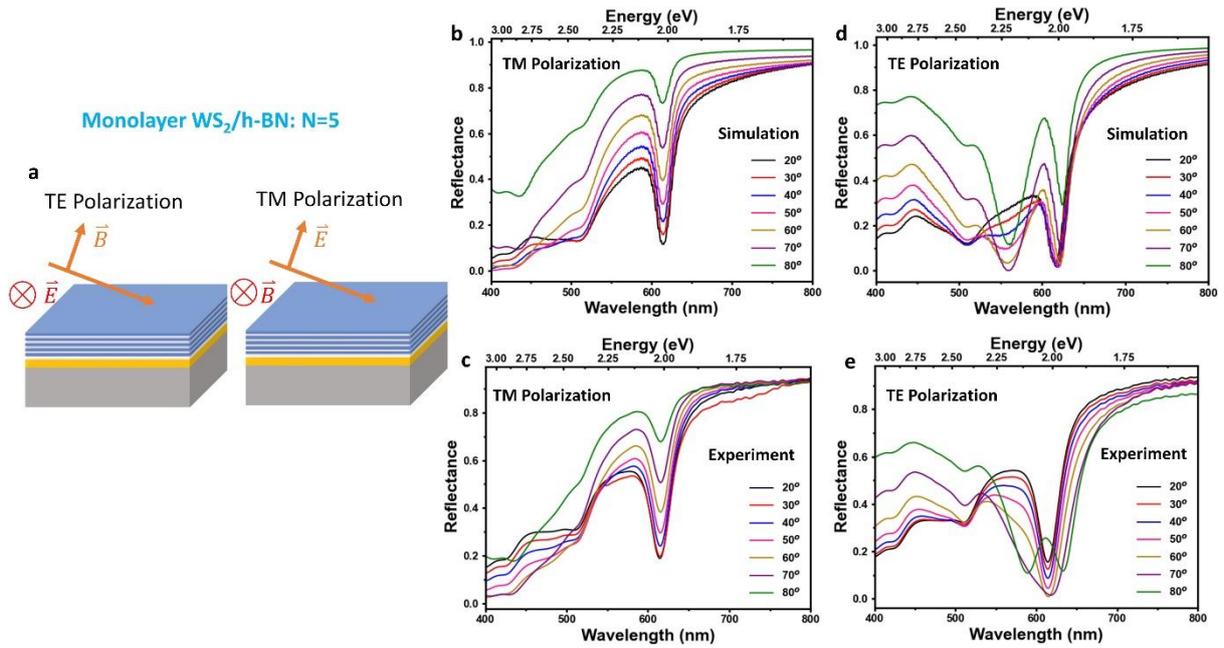

**Figure S19:** (a) depicts the two polarizations of incident light on the superlattice. For TE polarized light, the incident magnetic field tilts out of plane, and for TM polarized light, the incident electric field tilts out of plane. The (b) simulated and (c) experimental angle-resolved reflectance spectra of the $WS_2$/h-BN N=5 superlattice with TM polarized incident light. The (d) simulated and (e) experimental angle-resolved reflectance spectra of the $WS_2$/h-BN N=5 superlattice with TE polarized incident light.

The Poynting vector distribution of the cavity modes in $WS_2$ and $MoS_2$ superlattices are shown in Figure S20. The Poynting vector distribution was calculated using our transfer matrix method simulations. Since both of the superlattices have similar refractive index profiles owing to them have the same geometries. $WS_2$ and $MoS_2$ having similar refractive indices at λ=721 nm (n=4.04 for $WS_2$ and n=4.09 for $MoS_2$), both superlattices have the same cavity mode energy (1.72 eV). The Poynting vector of both superlattices was found to concentrate in the TMDC layers indicating that the increased absorption is due to waveguide modes in the TMDCs.

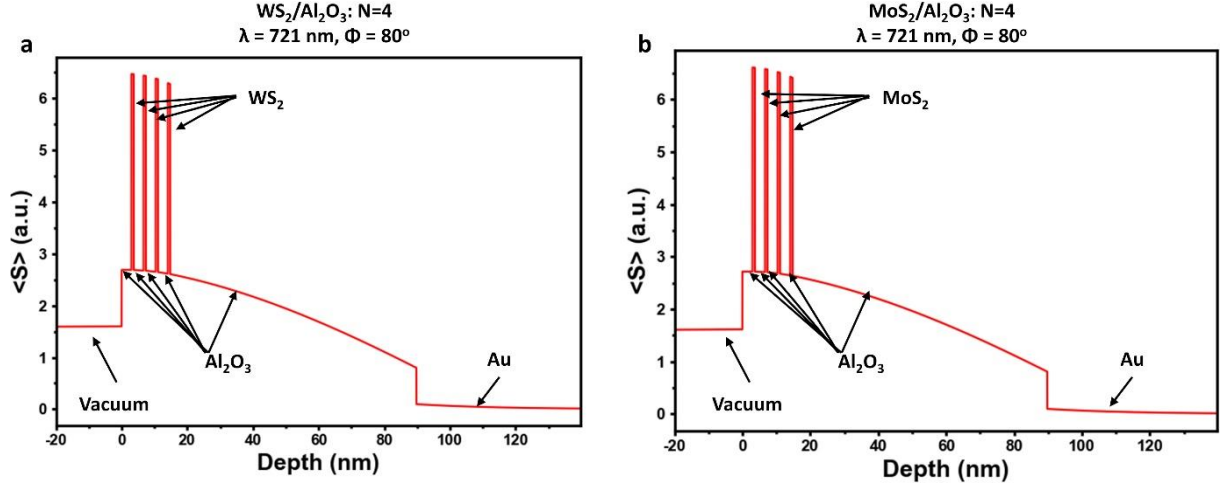

**Figure S20:** The Poynting vector distributions in (a) $WS_2/Al_2O_3$: N=4 and (b) $MoS_2/Al_2O_3$: N=4 superlattices. Both are for incident angles of 80° with bottom $Al_2O_3$ thicknesses of 75 nm. The arrows indicate the material in each region.

## Rabi splitting and energy calculation:

We modelled the exciton-polaritons in $WS_2$-based superlattices using a coupled oscillator model which is based on the Jaynes-Cummings Model Hamiltonian[3]

$$H = \begin{pmatrix} E_x & g \\ g & E_c(t_{Al2O3}) \end{pmatrix} \quad \ldots\ldots (1)$$

Where $E_x$ and $E_c$ are the uncoupled energies of the exciton and cavity modes, respectively, and g is the coupling parameter which is related to the Rabi splitting as $g = \frac{\hbar\Omega_{Rabi}}{2}$. $E_c$ was determined to be linearly dependent on the thickness of the bottom alumina layer, $t_{Al2O3}$. Assuming the damping factor of the cavity mode is much larger than the damping factor of the exciton, the splitting of the exciton-polaritons is in the strong coupling regime when $g > \frac{\gamma_C}{4}$ [4] where $\gamma_C$ is the damping factor of the cavity mode which is related to the Q factor of the mode by $Q = \frac{E_c}{\gamma_c}$. We found that our superlattices were all in the strong coupling regime as the incident angle approached 90°, and strong coupling occurred at lower incident angles as N increased (see Figure S24A).

The model was fitted to the simulated values using g and the linear dependence of the cavity mode on the bottom alumina thickness ($E_c$ = $mt_{Al2O3}$ + b) as the fit parameters and by using a least squares optimization method. Figure S19A shows that our model and the simulated energies of the UEP and LEP agree.

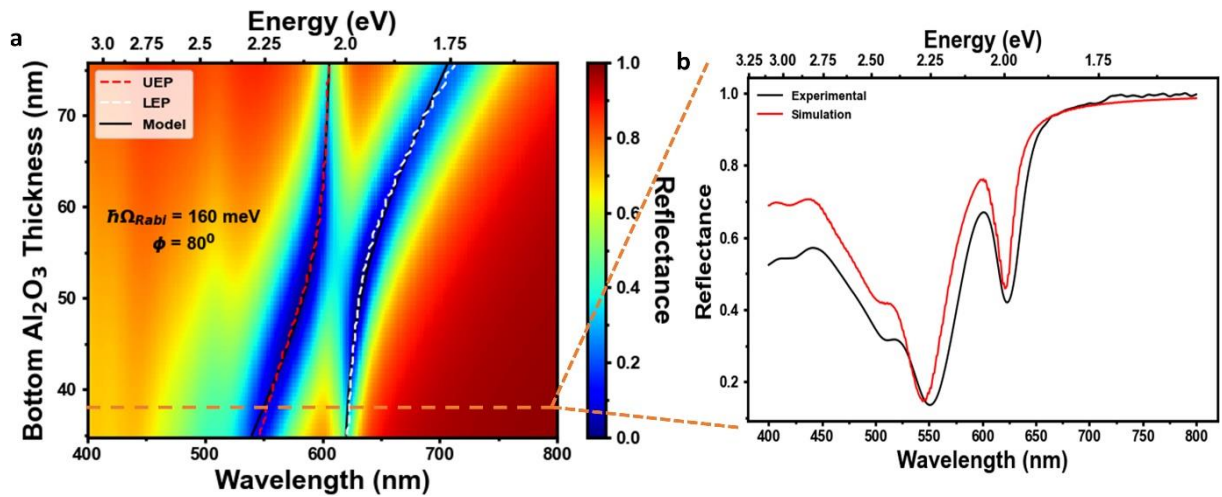

**Figure S21: Depicts the coupled-oscillator model used on the WS$_2$ based superlattices. (a)** Shows the simulated anti-crossing behavior of the WS$_2$/Al$_2$O$_3$ N=4. **(b)** Compared the simulated and experiment reflectance spectra of the WS$_2$/Al$_2$O$_3$ N=4 superlattice with a bottom alumina thickness of 34 nm.

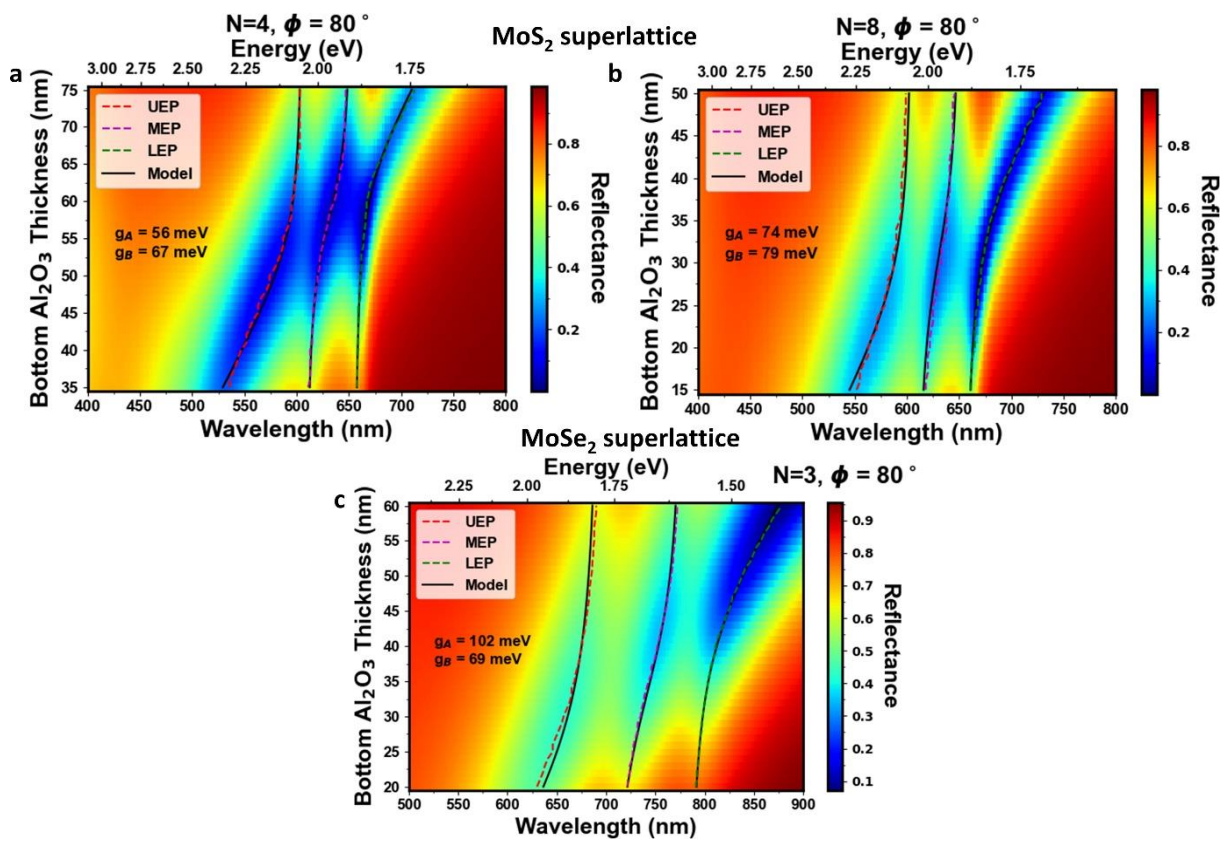

**Figure S22: The three-coupled oscillator model used for MoS$_2$/Al$_2$O$_3$ superlattices.** The simulated strong coupling behavior for MoS$_2$/Al$_2$O$_3$ superlattices with **(a)** N=4 and **(b)** N=8 unit cells. **(c)** Similar three coupled oscillator used in case of MoSe$_2$/Al$_2$O$_3$ superlattice with N= 3.

The energy difference between the A and B excitons in MoS$_2$ is smaller than in WS$_2$ allowing the cavity mode to interact with the A and B excitons simultaneously in MoS$_2$. This is reflected in the model as we used a three-coupled oscillator for MoS$_2$ where two of the oscillators are excitons while the third is the cavity. The Hamiltonian of this system can be written as[5]

$$H = \begin{pmatrix} E_A & 0 & g_A \\ 0 & E_B & g_B \\ g_A & g_B & E_C(t_{Al2O3}) \end{pmatrix} \quad \ldots\ldots (2)$$

Similar to the coupled oscillator model, the diagonal terms ($E_A$, $E_B$, and $E_C$) are the uncoupled energies of the A exciton, B exciton, and cavity mode, respectively, while the off-diagonal terms determine the coupling strength between oscillators. $g_A$ is the coupling parameter between the A exciton and the cavity mode while $g_B$ is the coupling parameter between the B exciton and the cavity mode. The 0 terms are due to the assumption that the A and B excitons do not couple with one another. This assumption was checked by allowing the terms to vary when fitting the model, but this approached gave the coupling between the excitons to be 5 orders of magnitude smaller than $g_A$ and $g_B$. The model was fitted to the simulated values using $g_A$, $g_B$, and the linear dependence of the cavity mode on the bottom alumina thickness as the fit parameters and a least square optimization approach.

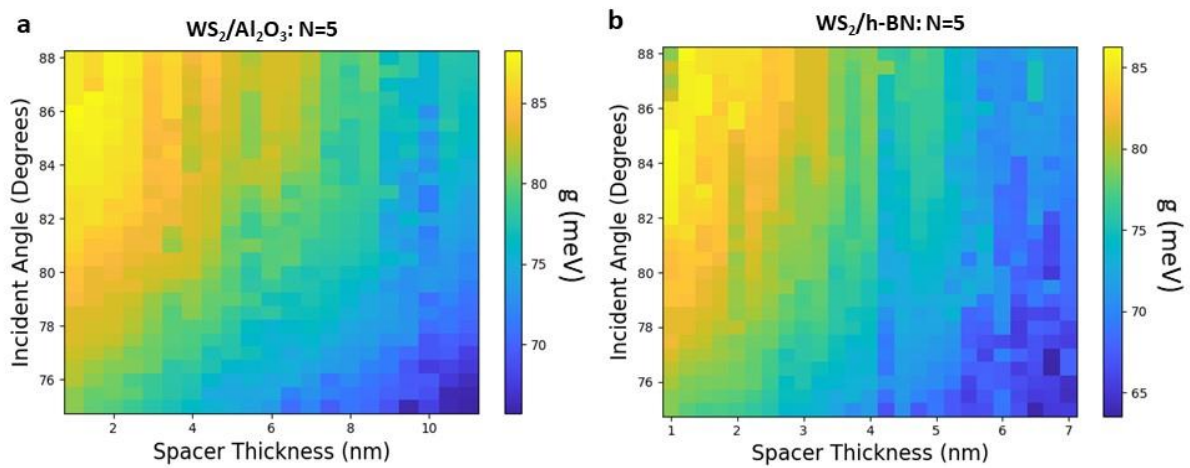

Figure S23: The angle and spacer thickness dependence on the coupling coefficient ($\frac{\hbar\Omega_{Rabi}}{2} = g$) for (a) WS$_2$/Al$_2$O$_3$ N=5 and (b) WS$_2$/h-BN N=5 superlattices.

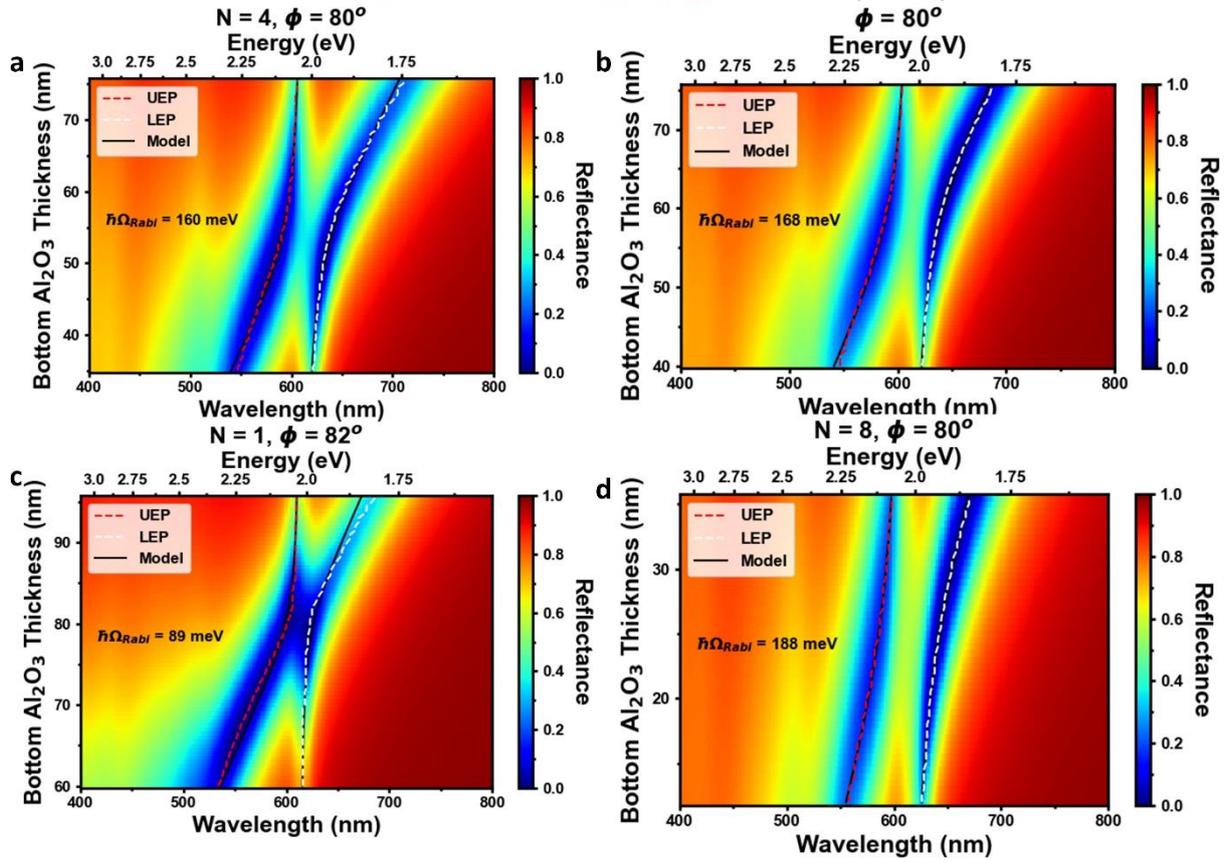

**Figure S24:** The simulated anti-crossing behavior of (a) $WS_2/Al_2O_3$ N=4, (b) bulk $WS_2$ (2.8 nm) on $Al_2O_3$ (34 nm) and Au, (c) $WS_2/Al_2O_3$ N=1, and (d) $WS_2/Al_2O_3$ N=8.

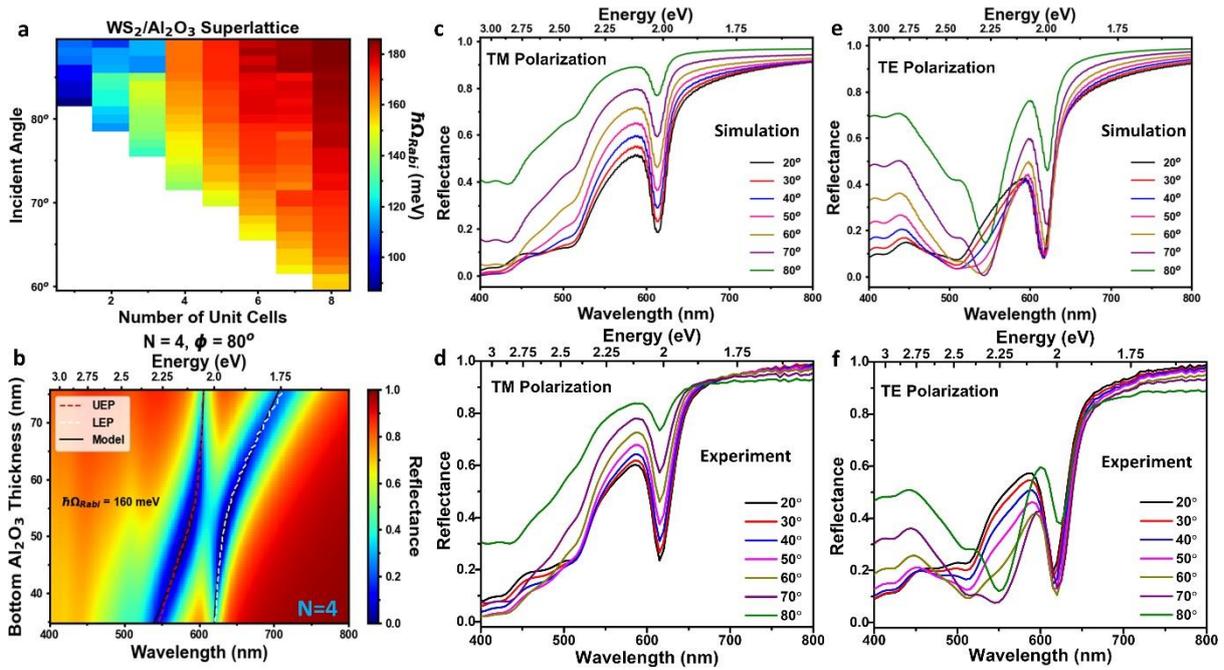

**Figure S25:** Spectra of the WS$_2$/Al$_2$O$_3$ N=4 superlattice. (a) The incident angle and number of unit cell dependence of the Rabi splitting under TE polarized light. (b) The anti-crossing behavior. The (c) simulated and (d) experimental reflectance spectra of TM polarized light. The (e) simulated and (f) experimental reflectance spectra of TE polarized light.

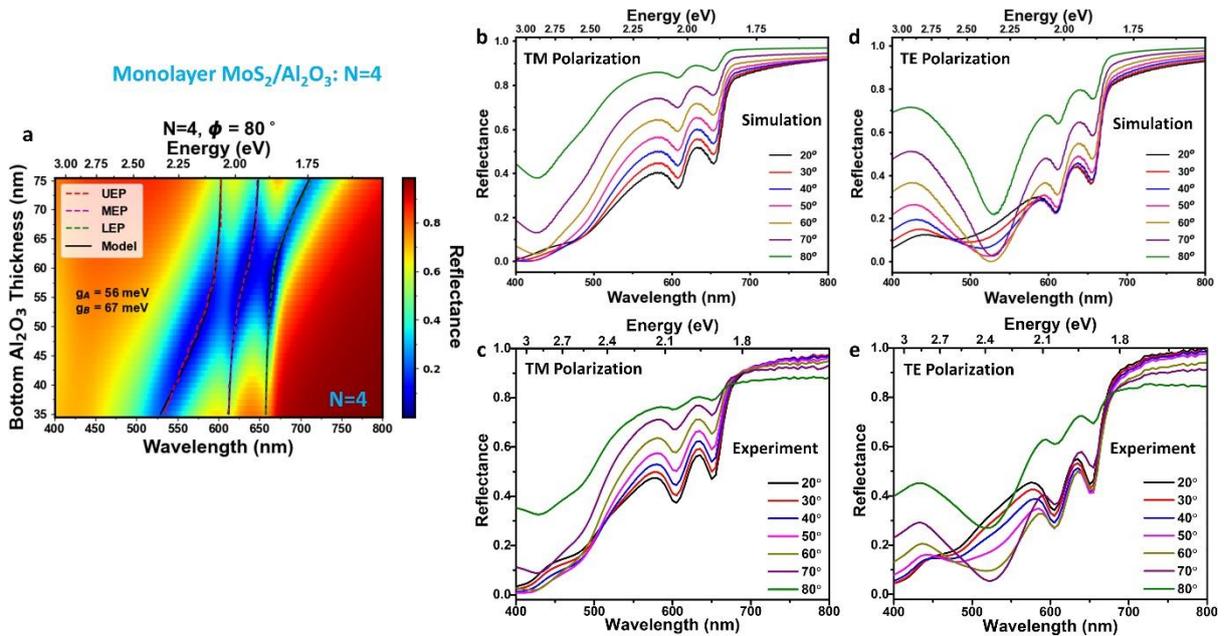

**Figure S26:** Spectra of the MoS$_2$/Al$_2$O$_3$ N=4 superlattice. (a) The anti-crossing behavior. The (b) simulated and (c) experimental reflectance spectra of TM polarized light. The (d) simulated and (e) experimental reflectance spectra of TE polarized light.

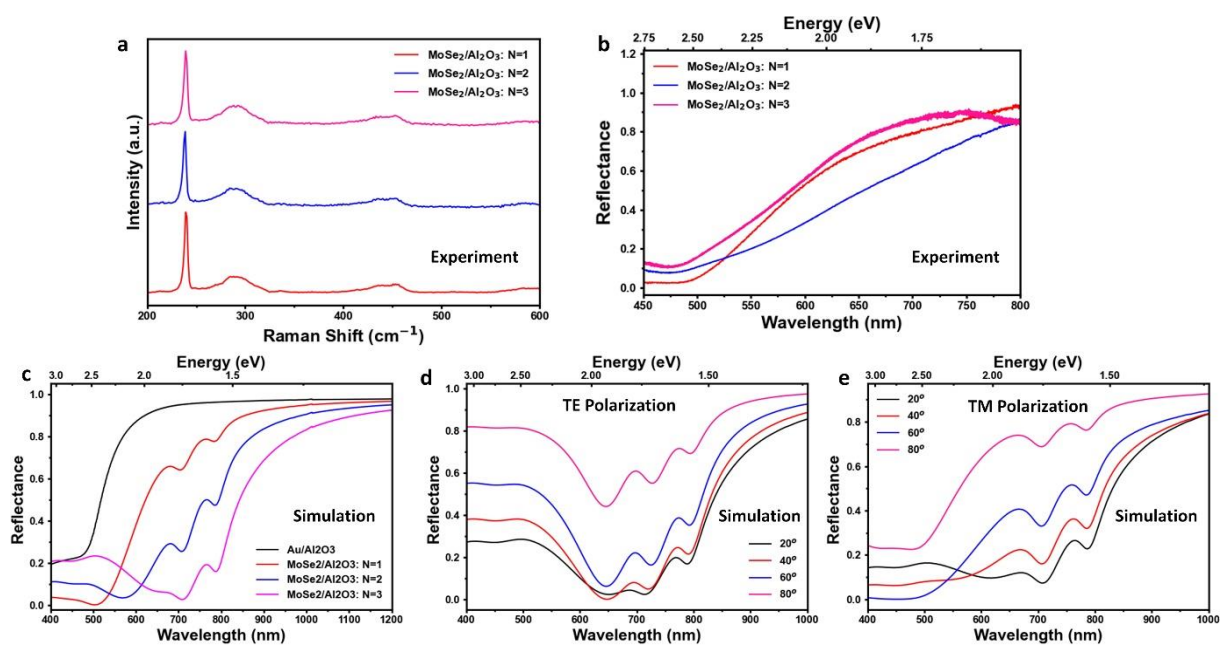

**Figure S27: The fabrication of MoSe$_2$/Al$_2$O$_3$ superlattice. (a, b) The layer dependent (unit cell N) Raman and reflectance spectra and the corresponding (c) simulated layer dependent reflectance spectrum. (d, e) The angle dependent reflectance spectra for the N=3 superlattice for TE and TM polarized incident light.**